\documentclass[%
reprint,
nofootinbib,
amsmath,amssymb,
aps
]{revtex4-2}

\usepackage{subcaption}
\usepackage{adjustbox}

\usepackage{graphicx}
\usepackage{dcolumn}
\usepackage{bm}
\usepackage{amsmath}
\usepackage{mathtools}

\usepackage{graphicx}
\usepackage{dcolumn}
\usepackage{bm}
\usepackage{amsmath}
\usepackage{mathtools}

\begin{document}

\title{High temperature modulations, meso-scale interactions and hyperscaling breakdown in Ising models with frustration: some insights from thermodynamic geometry }

\author{Soumen Khatua}
\email{soumenk.ph21.ph@nitp.ac.in }
\affiliation{National Institute of Technology Patna, Ashok Rajpath, 800005, Patna, India}
\author{Riekshika Sanwari}
\email{riekshikas.phd19.ph@nitp.ac.in}
\affiliation{Patna Women's College, Patna}
\author{Vikram Patil}
\email{vikrampatil@nitp.ac.in}
\affiliation{National Institute of Technology Patna, Ashok Rajpath, 800005, Patna, India}
\author{Anurag Sahay}
\email{anuragsahay@nitp.ac.in}
\affiliation{National Institute of Technology Patna, Ashok Rajpath, 800005, Patna, India}

\begin{abstract}
In this work we revisit the Axial Third Nearest Neighbour Ising (A3NNI) chain and examine in detail some aspects of its phase behaviour ensuing from competing interactions and resulting frustration. We probe the phase behaviour with two complimentary tools: a microscopic two-point correlation function $\mathcal{C}(n)$ which we carefully construct after appropriate spin transformations, and a macroscopic, thermodynamic curvature $R$ which we obtain from the free entropy. We report novel observations of phenomena such as hyperscaling breakdown and intermediate-ranged modulations among others. The zero field thermodynamic curvature $R_0$ is shown to systematically sub-divide the ground state phases into regions of attractive or repulsive effective interactions of varying strength. Furthermore, $R_0$ brings forth the significance of some third order moments in describing the effects of frustration, including the multiphase lines. Combined use of both the probes in the short-ranged modulated order regime confirms and further clarifies the discussion in [1] regarding the appropriate measure of high temperature correlation length in this regime. 

\end{abstract}

\pacs{Valid PACS appear here}
\maketitle

\section{Introduction}
\label{intro}

A wide class of crystalline systems such as polytypes, binary alloys, ferroelectrics and some magnetic materials are seen to exhibit phases in which either the stacking sequence of layers or some thermodynamic properties like magnetization, mass or charge density, etc , undergo a periodic modulation, with the period being commensurate or incommensurate with the underlying lattice structure, \cite{nato,neubert}. 

One of the earliest and the most well known model serving as a prototype for spatially modulated structures is the Axial Next Nearest Neighbour Ising (ANNNI) model, \cite{elliot,selke,steph,julia,lieb}. The three-dimensional model can be envisaged as Ising spins in $2d$ layers stacked along an axis such that there is ferromagnetic coupling between adjacent layers and anti-ferromagnetic coupling between alternate layers. Within the $2d$ layers the spins are always ordered ferrromagnetically so that it is the axial direction that remains of interest. 

The $3d$ model is characterised by the existence of a disordered (paramagnetic) phase, a ferromagnetic phase, a modulated-phase region, an antiphase phase, and a novel multicritical point called the Lifshitz point where the first three phases meet. The modulated-phase region presents an infinite number of commensurate structures  with the phase boundaries converging to a multiphase point at zero temperature where they become degenerate. The two dimensional model has a similar rich phase behaviour. While the one-dimensional case does not display any long range order at finite temperature, it nevertheless  shows the effects of frustration not only in its ground state phases and in the associated disorder point but also in its short-range modulations and the associated disorder lines at finite temperatures, \cite{steph}.

Extensions and modifications of the ANNNI model are aimed at achieving a more realistic modelling of specific physical systems and are also useful in checking for the robustness of the characteristic ANNNI phase behaviour against perturbations. Some of the variations include addition of  magnetic field, defects, impurities or vacancies, or the inclusion of a third neighbour interaction or competition along more than one axis, etc, \cite{selke,lieb}. It is worth mentioning that all these modifications exhibit rich ANNNI-like phase structures, with sequences of long wavelength modulated phases springing from multiphase points. Of interest in this work is the Axial Third Nearest Neighbour Ising (A3NNI) model first introduced in \cite{sby}. While it shares some of the its qualitative features  with that of the ANNNI model, the phase structure of the A3NNI model turns out to be richer and more varied, including a larger number of coexisting ground states and, hence, multiphase {\it{lines}} bordering them. In this work we add to the features of phase behaviour of the A3NNI chain through our novel observations of hyperscaling breakdown in the anti-ferromagnetic ground state and intermediate ranged modulations in a sub-region of the ferromagnetic ground state.

Several approaches including mean field approximations, Monte Carlo simulations, renormalization group studies, high temperature and low temperature series expansions, etc have been utilized in the investigation of the rich phase behaviour of the axial Ising models, \cite{selke}. In \cite{khatua1} we had initiated a thermodynamic geometry based approach to investigate the ANNNI model. Thermodynamic geometry, in Ruppeiner's entropy formalism, is a powerful framework for investigating the thermodynamics and phase behaviour of physical systems, \cite{rupporiginal,rupprev}. It has been extensively used to probe several systems such as lattice spin models, simple fluids, quantum gases, strongly interacting matter, etc \cite{sahay1}-\cite{castorina}. 

In this work we further develop and refine the TG based investigations of frustrated spin systems with our exploration of the the A3NNI chain, of which the standard ANNNI model turns out to be a special case. With the expanded scope of investigations our geometric insights gain in clarity, conceptual depth and perspective.  The state space scalar curvature of the Riemann manifold corresponding to the A3NNI chain is extensively employed to illuminate the phase behaviour and make predictions about the nature and strength of the underlying effective interactions. Furthermore, our construction of the two-point spin-spin correlation function reveals novel phase behaviour such as intermediate ranged modulation of spins in the ferromagnetic phase. In some places, such as in the short ranged modulation regime, the correlation function and the state space scalar curvature probes complement each other to support a robust understanding of the high temperature correlation length. At the same time, the invariant thermodynamic scalar curvature is able shed light on meso-scale interactions that remain unseen by the correlation function or the response functions.

The paper is broadly organized as follows. In section \ref{sec1} we start with the one-dimensional A3NNI Hamiltonian and then obtain its correlation function, phase structure and scaling behaviour. We then discuss in some detail the short-ranged and intermediate-ranged modulations in the ferromagnetic ground state of the parameter space. In section \ref{sec3}, after giving a quick introduction to TG, we describe the properties of the zero field scalar curvature $R_0$ for reflection symmetric Hamiltonians. In section \ref{sec4} we delve into the state space geometry of our model, relate it to its phase behaviour and obtain a geometric map of its underlying meso-scale interactions. Finally we summarize our work in the section \ref{sec5} and try to place it in perspective. The appendix reviews the standard transfer matrix based calculations involving the one-dimensional Ising model and brings out the relation between the magnetic susceptibility and the correlation length mentioned in section \ref{sec1}.

\section{One-dimensional A3NNI model }
\label{sec1}

The general $3d$ A3NNI  model comprises planes stacked along an axis  such that the axial couplings are $J_1$, $J_2$ and $J_3$ between adjacent(nearest-neighbour), next-nearest and next-to-next nearest layers. Within the layers the spin couplings are strongly ferromagnetic. For the one-dimensional case the layers are replaced by lattice points. The one-dimensional Hamiltonian in the presence of field is written as, \cite{sby},
\begin{eqnarray}
\mathcal{H}=&-&J_1\sum_{i}S_iS_{i+1}-J_2\sum_{i}S_iS_{i+2}\nonumber\\&-&J_3\sum_{i}S_iS_{i+3}-H\sum_i S_i
\label{hamiltonian}
\end{eqnarray}
 The ANNNI chain becomes a special case of the A3NNI with $J_3$ set to zero. We shall allow all the couplings $J_i$'s in the A3NNI Hamiltonian to range from positive to negative and also shall not place any restriction upon the relative strengths of the couplings. 
 
 While in this work we shall concern ourselves exclusively with the zero field axial Ising models it will be seen that a non-zero $H$ is necessary to set up a two-dimensional Riemannian state space of the system. Thereafter, by setting $H$ to zero one recovers the zero field thermodynamic curvature. In zero field one can see that the above Hamiltonian remains unchanged on simultaneously changing $\{J_1,J_3\}\to\{-J_1,-J_3\}$ and flipping odd spins, $S_i\to-S_i$, for $i$ odd. Therefore, since the phase structure for $J_1<0$ can be arrived at by simultaneously flipping the odd spins and the $J_3$ axis in the phase diagram of $J_1>0$  we may confine attention only to the case of positive $J_1$. We shall henceforth scale away $J_1$ and relabel parameters to 
\begin{equation}
\frac{1}{J_1}\{J_2,J_3,H,\mathcal{H}\}\to \{K,L,H,\mathcal{H}\}
\end{equation}  
 The zero field Hamiltonian can then be re-expressed as
  \begin{equation}
  \mathcal{H}=-\sum_{i}S_iS_{i+1}-K\sum_{i}S_iS_{i+2}-L\sum_{i}S_iS_{i+3}
  \label{hamiltonian zero}
  \end{equation}

  Starting with the full Hamiltonian in Eq.\ref{hamiltonian} the transfer matrix $\mathcal{U}$ can be written as
  \begin{eqnarray}
\langle u,v,w|\,\mathcal{U}\,|x,y,z\rangle&=&\exp\beta\left[\frac{1}{2}\,H\,(u+v+w+x+y+z)\right.\nonumber\\
  & +&  L\,(u\,x+v\,y+w\,z)
 +K\,(u\,w+v\,x\nonumber\\&+&w\,y+x\,z)
    +\frac{1}{2}\,(u\,v+v\,w+2w\,x\nonumber\\
   &+& \left.  x\,y+y\,z)\right]\nonumber\\
   \label{transfer matrix 1}
  \end{eqnarray}
  
  \begin{figure}[!t]
\centering
\includegraphics[width=2.8in,height=1.in]{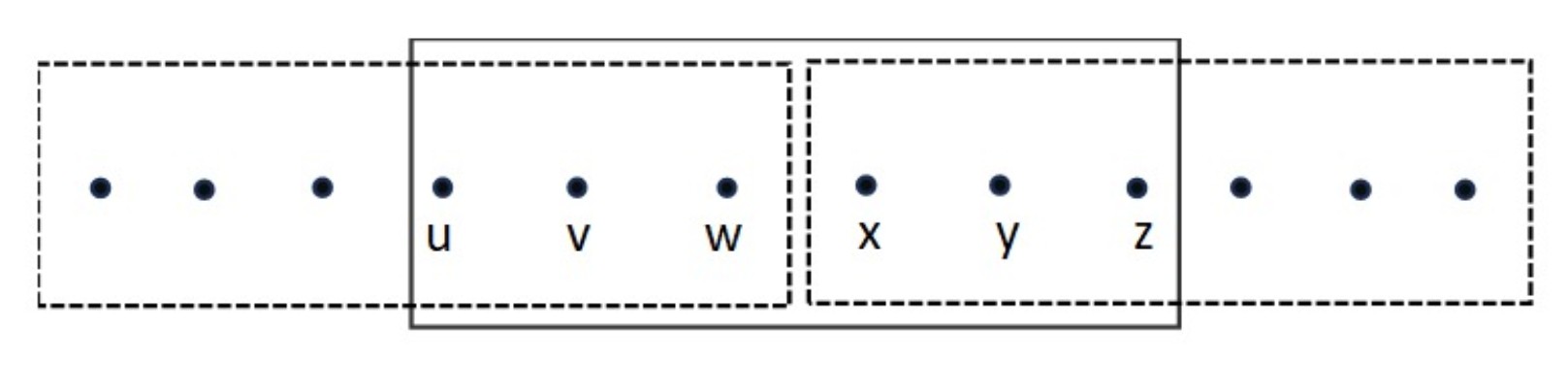}
\caption{\small{Block structure of the transfer matrix in Eq.\ref{transfer matrix 1} }}
\label{fig transfer}
\end{figure}
  where $u,v,..$ are spin labels with values $\pm 1$. 
  The transfer matrix, as shown in Fig.\ref{fig transfer} is an $8\times 8$ matrix, with each block Hamiltonian of $6$ lattice points and two blocks sharing $3$ Ising spins each so that there are $N/3$ blocks in total. Using the cyclic boundary condition the partition function $Z_N$ of $N$ lattice points can be expressed in the standard manner as the trace of $N/3$-th power of the transfer matrix. The eigenvalues can be obtained numerically and the logarithm of the largest eigenvalue $\lambda_+$ gives the Massieu function or the free entropy $\psi$ per lattice site as
  \begin{equation}
  \psi=\frac{1}{N}\log Z_N=\frac{1}{3}\log \lambda_+
  \label{massieu}
  \end{equation}
   where $\lambda_+$ is the largest eigenvalue of $\mathcal{U}$. Considering $\psi$ as a function of $\beta$ and $\nu =\beta H$ one can obtain useful thermodynamic quantities in the standard manner: the magnetization per lattice site $m=\partial\psi/\partial\nu$ and its fluctuation moment (variance) $\sigma_m^2=\partial^2\psi/\partial\nu^2$, the energy per lattice site $\epsilon=-\partial\psi/\partial\beta$ and its fluctuation moment per lattice site $\sigma_{\epsilon}^2=\partial^2\psi/\partial\beta^2$. The specific entropy $s$ is simply the Legendre transform $s=\psi-\beta\partial\psi/\partial\beta$.
   
   In fact, the eigenvalues $\lambda_i$'s can be obtained from a much simpler matrix $\mathcal{V}$,  
    \begin{equation}
 \langle u,v,w|\,\mathcal{V}\,|x,y,z\rangle= e^{u\,v + K\,u\,w+L\,u\,x+H\,u}\delta_{uy}\delta_{vz}
 \label{kassan ogly}
 \end{equation}
    This is the `unsymmetrized'  precursor to the transfer matrix as mentioned in Kramers and Wannier in the context of a nearest-neighbour Ising chain, \cite{kramers}. It was first used by Oguchi for third neighbour interactions, \cite{oguchi} and has been recently put to good use by Kassan-Ogly and co-workers for ANNNI and A3NNI chains in zero as well as non-zero magnetic fields, \cite{kassan}. While $\mathcal{V}$ and $\mathcal{U}$ have the same eigenvalues their eigenvectors differ.  $\mathcal{V}$ being a sparse matrix the numerical evaluation of eigenvalues becomes far more tractable compared to $\mathcal{V}$. It may be checked that the largest eigenvalue $\nu_+$ of $\mathcal{V}$ is related to $\lambda_+$ as
    \begin{equation}
    \nu_+=\lambda_{+}^{1/3}
    \label{lambda nu}
    \end{equation}
    so that we have $\psi=\log\nu_+$. We shall find it convenient to use this as the starting point for developing the thermodynamic geometry of the A3NNI chain.
   
\subsection{Correlation function}
   
  Following Dobson, \cite{dobson}, we obtain the zero field spin-spin correlation function by briefly changing from cyclic to open boundary condition, with the understanding that in the large $N$ limit the results of both the boundary condition equalize. We map to new spin variables $\sigma_i(i=0,\ldots,N-1)=\pm1$ such that
\begin{equation}
\sigma_0=S_1\,\,\,;\,\,\sigma_i=S_iS_{i+1}\,\,\,\,\,\,i=1\ldots(N-1)\nonumber
\end{equation}  

This transformation is invertible, with \begin{equation}S_i=\sigma_{i-1}\sigma_{i-2}\ldots\sigma_0,\nonumber\end{equation} so that the set of all $\sigma$-lattice configurations is the same as the set of $S$-lattice configurations.   
With the above mapping between $S$ and $\sigma$ it can be checked that the two-point correlation in the original $S$-lattice can be written as multiple correlations in the $\sigma$-lattice,
  \begin{equation}
  \langle S_iS_{i+r}\rangle=\langle\sigma_i\sigma_{i+1}\ldots\sigma_{i+r-1}\rangle
  \label{two point}
\end{equation}  

\begin{figure}[!h]
\centering
\includegraphics[width=2.8in,height=1.in]{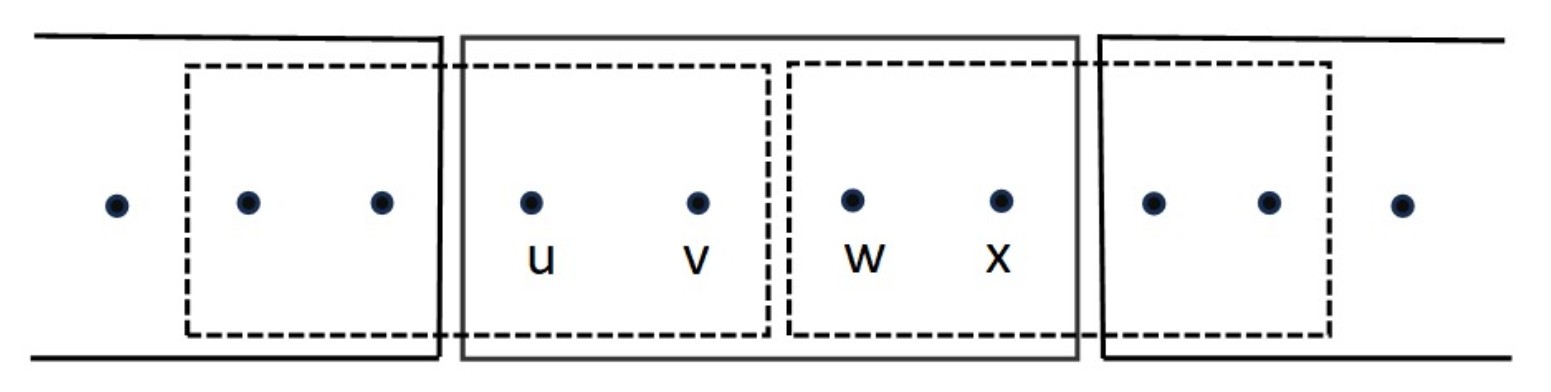}
\caption{\small{Block structure of the transfer matrix in Eq.\ref{transfer matrix 2}. }}
\label{fig sigma}
\end{figure}  

 In terms of the new spin variables the zero field Hamiltonian of Eq. \ref{hamiltonian zero} becomes
   \begin{equation}
   \mathcal{H}=-\sum_{i=1}^{N-1}\sigma_i-K\sum_{i=1}^{N-2}\sigma_i\sigma_{i+1}-L\sum_{i=1}^{N-3}\sigma_i\sigma_{i+1}\sigma_{i+2}
   \label{sigma hamiltonian}
   \end{equation}
Notice that $\sigma_0$ does not feature in the Hamiltonian.  The  partition function of the zero field Hamiltonian in the $\sigma$ representation becomes
 \begin{eqnarray}
 Z_N&=&\sum_{\sigma_0=-1}^{1}\sum_{\sigma_1=-1}^{1}\ldots\sum_{\sigma_N=-1}^{1} \exp \beta\mathcal{H}\nonumber\\
 &=& 2\,Q_{N-1}
 \label{partition sigma}
 \end{eqnarray}
 where the factor of $2$ arises due to a summation over $\sigma_0$. Therefore, the zero field thermodynamics can be equally computed from the partition function $Q_N$, which is expressed as the trace of $N/2$ multiples of the following $4\times 4$ transfer matrix,
 \begin{eqnarray}
 \langle u,v|\mathcal{T}|w,x\rangle &=& \exp\beta\left[\frac{1}{2}(u+v+w+x)+ \right.\nonumber\\
 &&\left. K(uw+vx)+L(uvw+vwx)\right]
 \label{transfer matrix 2}
\end{eqnarray}  
which represents a block of $4$ spins with $2$ spins shared by two blocks, as shown in Fig. \ref{fig sigma}.

  The realization of multiple correlations in the $\sigma$-lattice in Eq.\ref{two point} depends on whether the spin separation $r$ in the original lattice is even or odd. For even $r$ the multiple correlations in the $\sigma$-lattice can be obtained by inserting pairs of $\sigma_i\sigma_{i+1}$ between the transfer matrices. 
\begin{eqnarray}
&&  \langle S_iS_{i+r}\rangle= \langle S_1S_{1+r}\rangle=\langle\sigma_1\sigma_2\ldots\sigma_{r}\rangle =\nonumber\\&& \frac{1}{Q_{N-1}}\text{Tr}\left[\sigma_1\sigma_2\langle 1,2|T|3,4\rangle \sigma_3\sigma_4\langle 3,4|\mathcal{T}|5,6\rangle\ldots\right.\nonumber\\&&\left.\ldots\sigma_{r-1}\sigma_{r}\langle r-1,r|\mathcal{T}|r+1,r+2\rangle\ldots\right.\nonumber\\&&\left.\ldots\langle N-2,N-1|\mathcal{T}|1,2\rangle \right]\hspace{0.8in}\mbox{\small{(r even)}}
\label{multi correlation}
\end{eqnarray}
where we have reverted to cyclic boundary conditions and the `row' vector $\langle1,2|$ is $\langle\sigma_1\sigma_2|$, etc. Choosing an ordering so that $|\sigma,\sigma'\rangle$=$\{|++\rangle,|+-\rangle,|-+ \rangle,|-- \rangle\}$ are respectively the basis vectors $\{\begin{pmatrix}
1&0&0&0
\end{pmatrix}^T,\begin{pmatrix}
0&1&0&0
\end{pmatrix}^T,\begin{pmatrix}
0&0&1&0
\end{pmatrix}^T,\begin{pmatrix}
0&0&0&1
\end{pmatrix}^T\}$,
the combination $|\sigma,\sigma'\rangle\sigma\sigma'\langle\sigma,\sigma'|$ appearing in Eq.\ref{multi correlation} becomes the matrix ${\bf {P}}=\text{diag}[1,-1,-1,1]$
in terms of which the multiple correlation of Eq.\ref{multi correlation} can be written as
\begin{equation}
\langle S_1S_{1+r}\rangle=\frac{\text{Tr}\left(\tilde{\mathcal{T}}^{r/2} \mathcal{T}^{N-1-r/2}\right)}{\text{Tr}\left(\mathcal{T}^{N-1}\right)}\,\,\,\,\,\,(r\,\text{even})
\label{even}
\end{equation}
where $\tilde{\mathcal{T}}=\bf{P}\,\mathcal{T}$. For odd $r$ there will appear an unpaired spin $\sigma$ within the trace in Eq.{\ref{multi correlation} thus giving rise to a combination $|\sigma,\sigma'\rangle\sigma\langle\sigma,\sigma'|$  expressible as the matrix ${\bf {Q}}=\text{diag}[1,-1,1,-1]$. The spin-spin correlation for odd $r$ then becomes

\begin{equation}
\langle S_1S_{1+r}\rangle=\frac{\text{Tr}\left(\tilde{\mathcal{T}}^{(r-1)/2}\acute{\mathcal{T}} \mathcal{T}^{N-1-(r+1)/2}\right)}{\text{Tr}\left(\mathcal{T}^{N-1}\right)}\,\,\,\,\,\,(r\,\text{odd})
\label{odd}
\end{equation}
where $\acute{\mathcal{T}}={\bf {Q}}\mathcal{T}$.
As long as the eigenspace of the matrices $\mathcal{T}$, $\tilde{\mathcal{T}}$ and $\acute{\mathcal{T}}$ is four-dimensional, which we always find to be the case, we can diagonalize them with appropriate matrices constructed out of their respective eigenvectors, $\mathcal{T}={\bf{A}}\,{\bf{\Gamma}}\,{\bf{A}}^{-1}$, $\tilde{\mathcal{T}}={\bf{B}}\,{\bf{\tilde{\Gamma}}}\,{\bf{B}}^{-1}$ and  $\acute{\mathcal{T}}={\bf{C}}\,{\bf{\acute{\Gamma}}}\,{\bf{C}}^{-1}$ with $\bf{\Gamma}=\text{diag}\,[
\gamma_1,\gamma_2,\gamma_3,\gamma_4
]$, $\bf{\tilde{\Gamma}}=\text{diag}\,[
\tilde{\gamma}_1,\tilde{\gamma}_2,\tilde{\gamma}_3,\tilde{\gamma}_4
]$ and  $\bf{\acute{\Gamma}}=\text{diag}\,[
\acute{\gamma}_1,\acute{\gamma}_2,\acute{\gamma}_3,\acute{\gamma}_4
]$, the respective eigenvalues being arranged in a descending order. Note that following Perron-Frobenius theorem $\gamma_1$ is real and positive while there is no such condition on $\tilde{\gamma}_1$ or  $\acute{\gamma}_1$. It can be checked that $\gamma_1=\nu_+$ from Eq.{\ref{lambda nu} and hence $\gamma_1\geq|\tilde{\gamma}_1|,|\acute{\gamma}_1|$. The matrices ${\bf{A}}$,$\textbf{B}$ and ${\bf{C}}$ are complex in general.

Finally, in the limit of large $N$, the spin-spin correlation function can be written as

\begin{equation}\boxed{
\mathcal{C}(r)= (r\,\text{mod}\,2)\,\mathcal{C}_{odd}(r)+(1-r\,\text{mod}\,2)\,\mathcal{C}_{even}(r)
}\label{corr funct main}
\end{equation}

where
\begin{equation}
\mathcal{C}_{even}(r)=\frac{\text{Tr}\left({\bf{B}}^{-1}\,{\bf{A}}\,{\bf{\tilde{\Gamma}}}^{r/2}\,{\bf{A}}^{-1}\,{\bf{B}}\,{\bf{W}}\right)}{\gamma_1^{r/2}}
\end{equation}
and
\begin{equation}
\mathcal{C}_{odd}(r)=\frac{\text{Tr}\left({\bf{B}}^{-1}\,{\bf{A}}\,{\bf{\tilde{\Gamma}}}^{(r-1)/2}\,{\bf{A}}^{-1}\,{\bf{C}}\,{\bf{\acute{\Gamma}}}\,{\bf{C}}^{-1}\,{\bf{B}}\,{\bf{W}}\right)}{\gamma_1^{(r+1)/2}},
\end{equation}
with ${\bf{W}}=\text{diag}[1,0,0,0]$. Note that the above expression for correlation function in Eq.(\ref{corr funct main}) is different, and more general, than the one we presented in \cite{khatua1}. We shall find $\mathcal{C}(r)$ to be a useful probe of phase behaviour in the following. 

For a large spin separation $r$ the ratio $(\tilde{\gamma}_i/\gamma_1)^{r/2}$ is much larger for $i=1$ (and also  $i=2$ for complex conjugate $\tilde\gamma_{1,2}$) than for other $i$'s. Therefore, for large $r$, the correlation function goes as
\begin{equation}
\mathcal{C}(r)\sim e^{-r/\xi}\,\cos(q\,r)\,\,\,\,\,\,\,\text{(for large $r$)}
\label{corr large n}
\end{equation}
with the correlation length $\xi$ and the wavenumber $q$ given as
\begin{equation}
\xi^{-1}=\frac{1}{2}\log\frac{\gamma_1}{{|\tilde\gamma}_1|}\,\,\,;\,\,\,\,q=\arctan\frac{\text{Im}\,\tilde\gamma_1}{\text{Re}\,\tilde{\gamma_1}}
\label{corr}
\end{equation}
Complex $\tilde{\gamma}_1$ leads to an oscillatory correlation while real $\tilde\gamma_1$ leads to a monotonically decaying correlation.

 \subsection{Phase structure and scaling behaviour}
 \label{sub1}
 
 \begin{figure}[!t]
\centering
\includegraphics[width=3.2in,height=3.2in]{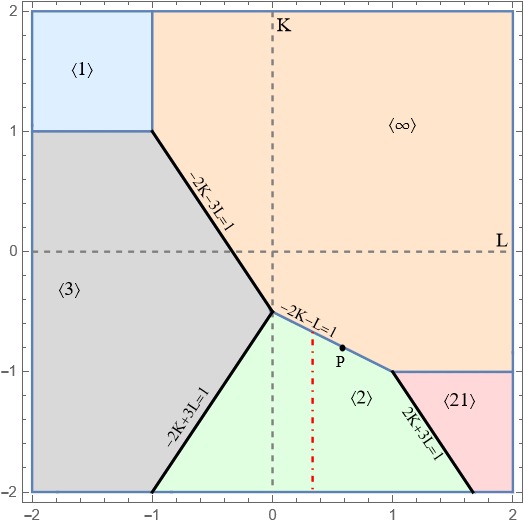}
\caption{\small{Ground state phase structure of the A3NNI chain in the $L$-$K$ parameter space. }}
\label{fig phase diagram}
\end{figure}
 
  The presence of higher order couplings allows for the possibility of a larger number of stable as well as frustrated spin configurations in the A3NNI model. Not surprisingly, owing to the extra $J_3$ coupling, the ground state phase structure of the A3NNI chain turns out to be richer than the ANNNI chain and has some features in common with the two-parameter ANNNI chain discussed in \cite{khatua1}. The ground state phase structure  may be obtained by carefully comparing the specific energy (energy per lattice point) of all the different allowed spin configurations. This process may be aided by an algorithmic process \cite{morita}.  A convenient way to represent repeating sequences of spin is the $m$-band representation of Selke and Fisher, \cite{selke}, where $\langle l_1l_2...l_m\rangle$ stands for a repetitive pattern of $m$ bands, with all spins in a given band of length $l_i$  ($i=1\cdots m$)  having the same orientation and adjacent bands being of opposite orientation. In this representation, for example, $\langle 2\rangle$ stands for the ground state $..\uparrow\uparrow\downarrow\downarrow\uparrow\uparrow\downarrow\downarrow...$ while $\langle 23\rangle$ labels $...\uparrow\uparrow\downarrow\downarrow\downarrow\uparrow\uparrow\downarrow\downarrow\downarrow...$ or its mirror image $...\downarrow\downarrow\uparrow\uparrow\uparrow\downarrow\downarrow\uparrow\uparrow\uparrow...$ depending on whether the field goes to zero from the negative or the positive side.

     In Fig. \ref{fig phase diagram}  we plot the ground state phase diagram (first reported in \cite{sby}) of the A3NNI model in the $L$-$K$ plane in the limit $H\to +0$. The phase diagram shows the coexistence of five phases,  namely ferromagnetic or $\langle\infty\rangle$, antiferromagnet $\langle 1\rangle$, the two antiphase configurations $\langle 2\rangle$ and $\langle 3\rangle$, and the ferrimagnet $\langle 21\rangle$.   The finite temperature wavelength\footnote{Not to be confused with the eigenvalue $\lambda_+$ in Eq.\ref{massieu}.} $\lambda$ of oscillations in the anti-phase regions is given as $\lambda=2\pi/q$ according to Eq.(\ref{corr}) with the exception of the $\langle21\rangle$ region where $\lambda=\pi/q$. This is because its period of $3$ is odd so that, unlike the $\langle 3\rangle$ and $\langle 2\rangle$ phases, its $\mathcal{C}_{even}$ or $\mathcal{C}_{odd}$ will repeat at $2\times 3=6$ lattice points. 
     
   We label the phase boundaries, seven in number, using a simple notation: $\langle\infty|1\rangle,\langle\infty|3\rangle,\langle\infty|2\rangle,\langle\infty|21\rangle$, and $\langle1|3\rangle,\langle 3|2\rangle,\langle2|21\rangle$. Thus, $\langle\infty|1\rangle$ separates the ferromagnetic and the antiferromagnetic ground states, etc. Note that the dashed red coloured vertical line at $L=1/3$ in the $\langle 2\rangle$ region of Fig.\ref{fig phase diagram} is not a phase boundary. We will describe it shortly. 
   
   As is well known, a few of the phase boundaries are lines of frustration containing infinitely degenerate ground state configurations, with the resulting specific entropy being non-zero along these lines. These lines are $\langle\infty|3\rangle$ with specific entropy (in units of $k_B$) of $0.382$ and $\langle 3|2\rangle,\langle 2|21\rangle$ each with specific entropy $0.281$. They are drawn as thick lines in Fig.\ref{fig phase diagram}. The triple point lying on the $K$-axis at the intersection of lines $\langle 2|3	\rangle$, $\langle\infty|3\rangle$ and $\langle\infty|2\rangle$ is maximally frustrated, with an entropy $(1+\sqrt{5})/2=1.618$. While we have obtained the entropies directly from the partition function the same may be obtained more elegantly by counting all possible degenerate configurations, \cite{milos}. In higher dimensional A3NNI lattices these frustrated lines are appropriately termed multiphase lines into which multiple phase coexistence lines converge at zero temperature, \cite{sby}. From these multiplhase lines spring forth a multitude of commensurate phases with a wide range of progressively complex periodicities (see Fig.\ref{fig multiphase}). It will be interesting to see how these degenerate boundaries are represented in the state space geometry.
   
\begin{figure}[h]
\centering
\includegraphics[width=3in,height=3in]{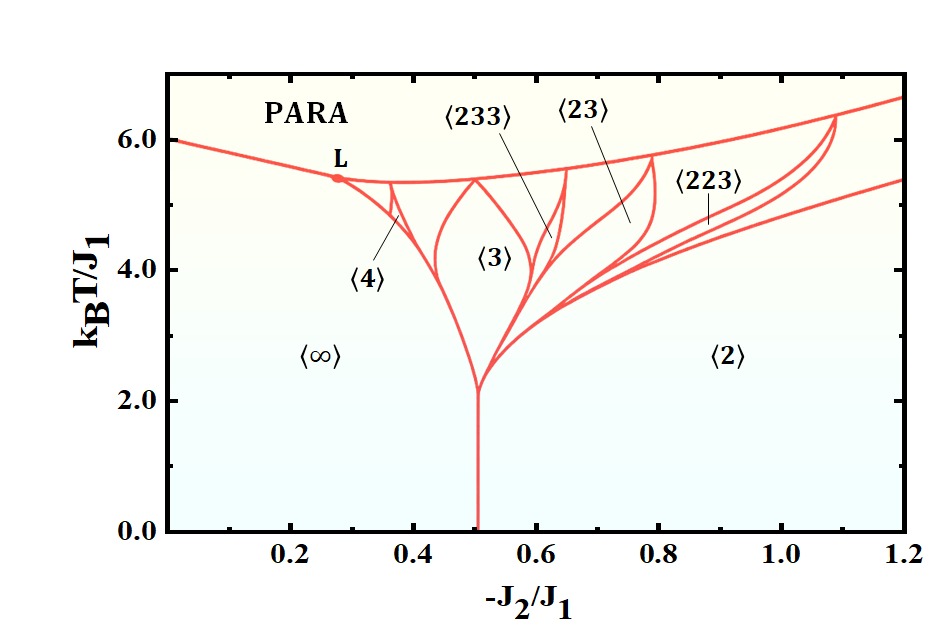}
\caption{\small{Schematic, a mean field phase diagram showing the convergence of several commensurate phases (only a few are labeled) into the multiphase point at $J_2/J_1=-0.5$. Adapted from figure 4 of \cite{sby}. The point marked $L$ is the Lifshitz point where three phases meet. Here $J_3/J_1=0.01$}}
\label{fig multiphase}
\end{figure}

\begin{table*}[t!]  
  \centering
 \scalebox{1.3}{
   \begin{tabular}{|c|c|c|c|}
\hline \text { Region } &   $\psi_s$ &  $\xi$ & $\sigma_m^2$ \\
\hline$\langle\infty\rangle$ & $e^{-2 \beta(J+2 K+3 L)}$ & $\frac{1}{2} e^{2 \beta(J+3 L+2 K)}$ &  $e^{2 \beta(J+3 L+2 K)}$ \\

\hline$\langle 3\rangle$ & $\frac{1}{3} e^{-\beta\left(-\frac{2 J}{3}-\frac{9 K}{3}-2 L\right)}$ & $6 e^{\beta\left(-\frac{2 J}{3}-\frac{4 K}{3}-2 L\right)} $& $\frac{1}{3} e^{-\beta\left(-\frac{2 J}{3}-\frac{4 K}{3}-2 L\right)}$ \\

\hline$\langle 3 | 1\rangle$ & $\frac{1}{3} e^{2 \beta(L+J)}$ & $6 e^{-2 \beta(L+J)}$ & $\frac{1}{3} e^{2 \beta(L+J)}$ \\

\hline$\langle 1\rangle$ & $e^{\beta(4 L-4 K+4 J)}$ & $\frac{1}{2} e^{\beta(-6 L+4 K-6 J)}$   & $4 e^{\beta(4 L-4 K+4 J)}$\\

\hline$\langle\infty| 1 \rangle$ & $e^{-2 \beta(J+3 L+2 K)}$ & $e^{ 2\beta(J+3 L+2 K)}$  & $e^{2 \beta(J+3 L+2 K)}$\\

\hline$\langle 2\rangle_L$ & $\frac{1}{2} e^{\beta(J+2 K-3 L)}$ & $2 e^{\beta(-J-2 K+3 L)}$ & $\frac{1}{2} e^{\beta(J+2 K-3 L)}$ \\

\hline$\langle 2\rangle_R$ & $\frac{1}{2} e^{-\beta(J-3 L-2 K)}$ & $2 e^{\beta(J-3 L-2 K)}$  & $\frac{1}{2} e^{-\beta(J-3 L-2 K)}$ \\

\hline$\langle\infty \mid 2\rangle_1$ & $\frac{7}{10} e^{-2 L \beta}$ & $\frac{10}{7} e^{2 L \beta}$ & $\frac{10}{7} e^{2 L \beta}$  \\

\hline$\langle\infty \mid 2\rangle_2$ & $\frac{1}{2} e^{-\beta(-2 L+2J)}$ & $2 e^{\beta(-2 L+2J)}$  & $\frac{1}{2} e^{-\beta(-2 L+2J)}$\\

\hline$\langle\infty \mid 21\rangle$ & $\frac{1}{3} e^{-\beta(2 L-2 J)}$ & $6 e^{\beta(2 L-2 J)}$ & $\frac{1}{3} e^{\beta(2L-2 J)}$\\

\hline$\langle 21\rangle$ & $\frac{1}{3} e^{-\beta\left(\frac{4 K}{3}+2 L-\frac{2 J}{3}\right)}$ & $6 e^{\beta\left(\frac{4 K}{3}+2 L-\frac{2 J}{3}\right)}$ & $\frac{1}{3} e^{\beta\left(\frac{4 K}{3}+2 L-\frac{2 J}{3}\right)}$\\
\hline

\end{tabular}
}
  \caption{\small{Asymptotic form of the singular free energy $\psi_s$, the correlation length $\xi$ and the variance in magnetic fluctuations $\sigma_m^2$ in different regions. }}
  \label{tab psicorr scaling}
\end{table*}

In Table \ref{tab psicorr scaling} we note our observations of the scaling behaviour of the (singular) free entropy $\psi_s$, the correlation length $\xi$ and the magnetic fluctuations $\sigma_m^2$ near the ground state in different regions of the parameter space. The singular free entropy is obtained everywhere by filtering out the non-singular, linear part of the Massieu function $\psi$ obtained in Eq.(\ref{massieu}) and retaining the slowest decaying terms as the temperature goes to zero. The asymptotic behaviours of both $\psi_s$ and $\xi$ are obtained by numerical considerations. 

With the exception of the anti-ferromagnetic phase (region $\langle 1\rangle$ in Fig.\ref{fig phase diagram}) we see from the table that hyperscaling is maintained in all the ordered regions, with singular free entropy per lattice $\psi_s$ approaching zero and the correlation length diverging towards the ground state in such a manner that,
\begin{equation}
\psi_s\,\xi\sim \mathcal{O}(1)\,\,\,\,\mbox{(hyperscaling near criticality)}
\label{hyperscaling}
\end{equation}
 This is consistent with the behaviour near criticality where the correlation 
volume ( $\xi$ in one-dimension) behaves as a single effective degree of freedom with cooperative microscopic fluctuations across its size, \cite{pathria}. The product of the singular free entropy per lattice site $\psi_s$ and the correlation length $\xi$ is therefore the entropy associated with a single effective degree of freedom of the correlated domain which is correctly of order unity. 

Here it will be useful to point out that for our one-dimensional lattice which undergoes pseudo-criticality at $T=0$ the singular free entropy per latice site $\psi_s$ is a better measure of hyperscaling that the more commonly used singular free \it{energy}} $f_s =-\psi_s/\beta$. In the latter case the product $f_s\,\xi$  would approach zero at the pseudo-critical point. That is, while the energy cost of fluctuating the single effective degree of freedom associated with the correlation domain goes to zero towards the ground state the entropy associated with it remains order unity.

Surprisingly, we find that hyperscaling is not maintained in the anti-ferromagnetic or the $\langle 1\rangle$ phase, with the correlation length there growing faster than the singular free entropy by the factor $e^{-2\beta(L+J)}$. That is, we have that
\begin{equation}
\psi_s\,\xi\to\infty\,\,\,\,\,\,\mbox{(non-hyperscaling)}
\end{equation}
In other words the entropy associated with a correlated domain {\it{grows}} on approaching the ground state (though not as fast as $\xi$) thus implying that in this case we cannot associate only a single effective degree of freedom with the domain. 
Incidentally, in our earlier study in \cite{soum} of the spin-3/2 chain with a Blume-Emery-Grifiths Hamiltonian we had seen hyperscaling violation in some regions of the parameter space and had found that earlier Monte Carlo studies of the $2d$ model had reported violations in similar regions. It would be therefore interesting to check for this feature in higher dimensional A3NNI models. 

Further, we see that the anti-phase region $\langle 2\rangle$ of Fig.\ref{fig phase diagram} exhibits different scaling behaviours in two sub-regions separated by the vertical dashed red line at $L=J/3$. We also find that the boundary $\langle \infty|2\rangle$ splits at the point $P$ into two segments with different scaling. The line segment to the left of $P$ shows ordering in continuation with the $\langle\infty\rangle$ phase above while to its right it exhibits ordering in continuation with the $\langle 2 \rangle$ phase below.

Finally, we note that wherever $\sigma_m^2$ diverges (namely in the ferro- and ferrimagnetic regions and some associated boundaries) it asymtotically approaches $\xi$. On the other hand, in $\langle2\rangle$ and $\langle3\rangle$ where it decays, $\sigma_m^2$ approaches $\xi^{-1}$. The the obvious exception is the hyperscaling violating $\langle 1\rangle$ region where it continues to follow $\psi_s$, as in the rest of the anti-phase regions, but not $\xi^{-1}$. The relation of $\sigma_m^2$ to $\xi$ appears to be similar to the case of the Ising ferromagnet and anti-ferromagnet even though the structure of spin-spin correlations is quite different. This does suggest that the (inverse of) magnetic susceptibility $\chi_m=\beta\sigma_m^2$ may be effectively used to quantify critical behaviour even in cases of anti-phase ordering. The alternative, staggered magnetic susceptibility $\chi_{stag}$, is challenging to measure in a laboratory set up, as is well known. In the Appendix  we work out the relation between the magnetic fluctuations and the correlation length for the Ising chain. 

\subsection{Modulated order in the ferromagnetic phase}

Similar to the case of the canonical ANNNI model or the two-parameter ANNNI model there exist parts of the ferromagnetic parameter space in the A3NNI chain  where on increasing the temperature (i.e, lowering $\beta$) the decay of spin correlations transitions from monotonic to oscillatory. This occurs in the  sub-region of the ferromagnetic or the $\langle\infty\rangle$ phase where at least one of the couplings $K$ or $L$ is negative (see Fig.(\ref{fig phase diagram})). While the stable low temperature (ground state) configuration of this sub-region remains ferromagnetic, at higher temperatures entropy helps stabilise modulated configurations mediated by negative $K$ or $L$.

Unlike the earlier instances, however, the A3NNI case shows a variation in the nature of modulation across the parameter range. It turns out, the short ranged modulations with a temperature dependent wavelength, akin to the aforementioned ANNNI cases, occur only in the sub-region of the ferromagnetic parameter space for which the third nearest neighbour coupling $L$ remains negative. This sub-region borders with the $\langle1\rangle$ and the $\langle 3\rangle$ phases.  On the other hand, the positive $L$ sub-region adjoining the $\langle 2\rangle$ and the $\langle 21\rangle$ phases exhibits intermediate ranged modulations at high temperatures, with a wavelength that is temperature independent. For both the cases the modulation effects are stronger and persist upto lower temperatures the closer one gets to their respective adjoining phase boundaries and gradually weaken on moving away towards the first quadrant, with $K,L>0$.

\subsubsection{Short ranged modulations: range of order}
\label{sub1sub1}

 In this regime the wavelength of oscillatory decay of correlations is temperature dependent and diverges to infinity as the inverse temperature $\beta$ is raised to a threshold value $\beta_D$, beyond which the decay becomes monotonic. This temperature $\beta_D$ has been termed the {\textit{disorder point of the first kind}} in earlier works, \cite{steph,selke,lieb}. In Fig.\ref{fig contour plot disorder lines} we draw contour plots of $\beta_D$ in the short-ranged modulated order regime of the ferromagnetic phase. The contour lines of higher $\beta_D$ have larger negative values of $L$ for larger positive value of $K$. The higher $\beta_D$ contour lines are seen to align more and more with the $\langle\infty|3\rangle$ boundary.

     \begin{figure}[!t]
\centering
\includegraphics[width=3in,height=2.5in]{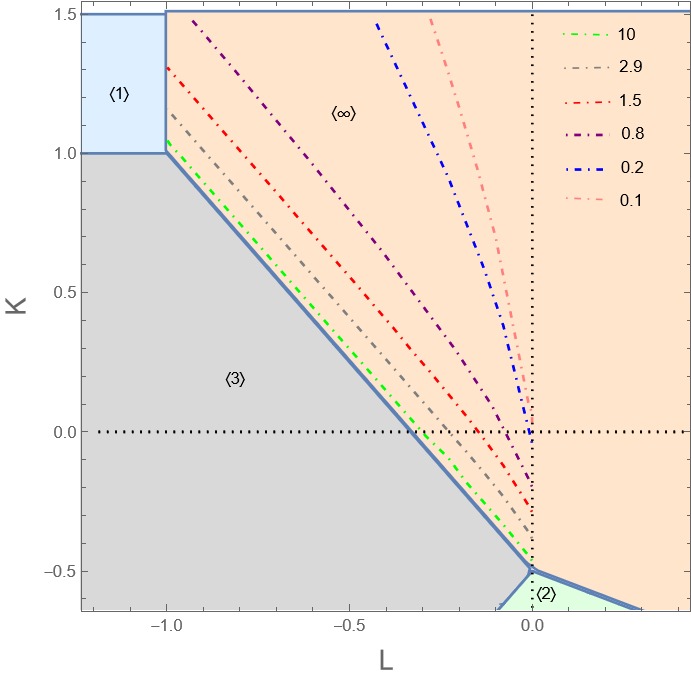}
\caption{\small{Contour plot in the $L$-$K$ parameter space of the disorder points $\beta_D$ in the short-ranged modulation regime of the ferromagnetic phase, i.e for $L<0$. }}
\label{fig contour plot disorder lines}
\end{figure}

 \begin{figure*}[htbp]
    \centering
    \begin{subfigure}{0.4\textwidth}
        \centering
        \includegraphics[width=\linewidth]{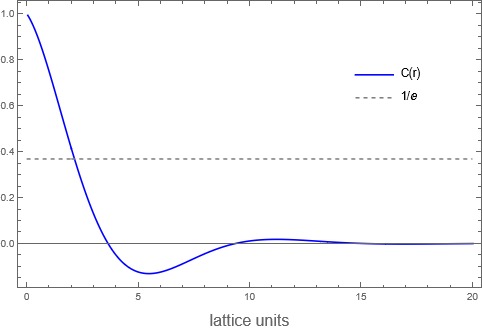}
        \caption{}
        \label{fig:ro1}
    \end{subfigure}%
    \hfill
    \begin{subfigure}{0.4\textwidth}
        \centering
        \includegraphics[width=\linewidth]{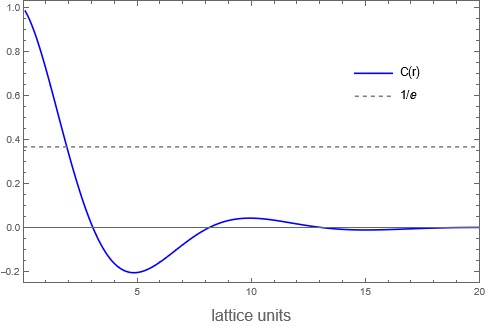}
        \caption{}
        \label{fig:ro2}
    \end{subfigure}%
   \\%
    \begin{subfigure}{0.4\textwidth}
        \centering
        \includegraphics[width=\linewidth]{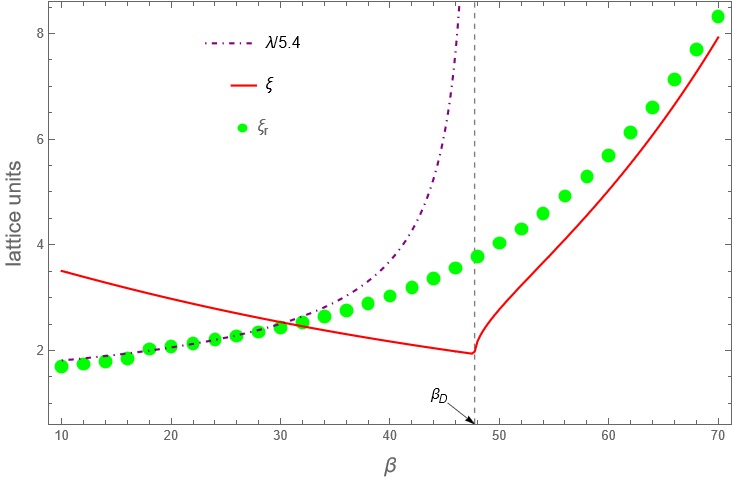}
        \caption{}
        \label{fig:ro3}
    \end{subfigure}%
    \hfill
    \begin{subfigure}{0.4\textwidth}
        \centering
        \includegraphics[width=\linewidth]{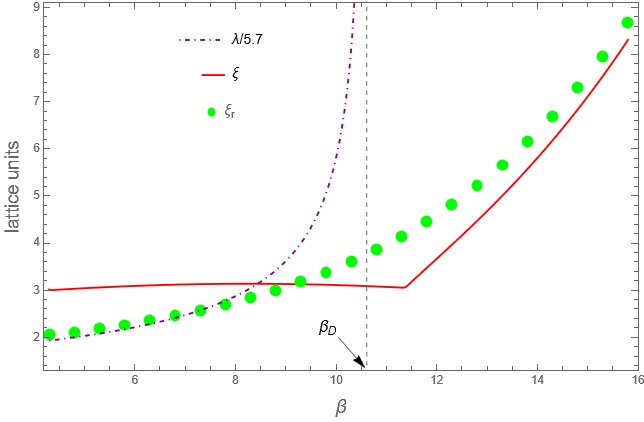}
        \caption{}
        \label{fig:ro4}
    \end{subfigure}%
    \caption{\small{($a$) and ($b$) show plots of the correlation function $\mathcal{C}(n)$ and $1/e$ $vs.$ the lattice units in  the sub-regions showing short-ranged modulations. The parameter values are $L=-0.166,K=-0.241$ and $\beta=22<\beta_D=47.74$ in ($a$) and $L=-0.99,K=1.03$ and $\beta=3<\beta_D=10.61$ in ($b$). The range of order $\xi_r$ is defined to be the point of intersection of the correlation function curve with the line $1/e$. Respective to the parameter values in ($a$) and ($b$) the sub-figures ($c$) and ($d$) show plots of the standard correlation length $\xi$, range of order $\xi_r$ and about a fifth of the modulation wavelength $\lambda$ with $\beta$ ranging across the disorder point $\beta_D$. See text for discussion.  }}
    \label{fig range of order}
\end{figure*}

In \cite{khatua1} we had pointed out what be believe to be an incorrect representation in earlier literature of the correlation length in the short-range modulated region. Obtained in the standard manner as the ratio of eigenvalues, see Eq.(\ref{corr}), $\xi$ was shown to decrease on lowering the temperature (or increasing $\beta$) till it reached a minimum at the disorder point $\beta_D$ where it underwent an abrupt upturn (discontinuous first derivative) and thereafter diverged in the usual manner at low temperatures (large $\beta$). Refer to the red curve in Fig.\ref{fig:ro3}. Mathematically, this is caused by the eigenvalue $\tilde{\gamma}_1$ in Eq.(\ref{corr}) changing across $\beta_D$ from complex to real resulting in the decay of correlations to change from oscillatory to monotonic. We had pointed out a subtle issue concerning the limits of validity of the standard expression for correlation length, especially in cases of rapid oscillatory decay, which was perhaps overlooked earlier. In the following we revisit the arguments, review our proposed resolution and also add some fresh perspective as well as observations.

 As implied earlier, the asymptotic expression for $ \mathcal{C}(r)$ in Eq.\ref{corr large n}, and hence of the correlation length $\xi$ in Eq.\ref{corr}, is strictly valid only when the spin separation $r$ is sufficiently large for the character of exponential decay of the correlation function to have been well established. While it is easy to judge monotonic exponential decay from the shape of the correlation function, for oscillatory decay one typically deals with the envelope of oscillatory decay in $C(r)$. At the same time it is also generally true that it is at lower temperatures (large $\beta$) that the correlation function remains finite for large $r$ whereas for higher temperatures (small $\beta$) it often diminishes too rapidly for it to be of any consequence at larger values of $r$. Thus at high temperatures due care must be taken in utilizing the asymptotic expression in Eq.\ref{corr}  for $\xi$.  This is especially true for the case of rapid oscillatory decay, since in such a scenario the `envelope' tends to quickly lose any meaning. 
 
 In such circumstances as discussed above we had found it illuminating to make direct use of the correlation function, instead of its envelope, to estimate the correlation length for short-ranged modulations. We termed our estimate  the `range of order'\footnote{The phrase `range of order' has been borrowed from Stephenson's work on ANNNI chain, \cite{steph}, where he had meant it to be the standard correlation length.} $\xi_r$ which we presently explain below.

 In Fig.\ref{fig:ro1} and Fig.\ref{fig:ro2} we show, respectively, for $K,L$ both negative and for $K$ positive, $L$ negative (at a point close to the anti-ferromagnetic border $\langle\infty|1\rangle$), the rapid oscillatory decay of the correlation function $\mathcal{C}(r)$. Clearly, the oscillatory decay of correlations in both the cases is too rapid and abrupt to allow for any meaningful interpretation of the `envelope' of decay. Instead we shall choose our correlation length as simply the point on the curve where the correlation drops to $1/e$. Thus, in the plots, the lattice distance at which the horizontal line $y=1/e$ intersects with the curve $y=\mathcal{C}(n)$ is our `range of order' $\xi_r$.

 In Figs.\ref{fig:ro3} and \ref{fig:ro4} we plot the standard correlation length $\xi$, the wavelength $\lambda=2\pi/q$ according to Eq. (\ref{corr})  and the range of order $\xi_r$ across the disorder point $\beta_D$ for the same $K,L$ values as in, respectively, Fig.\ref{fig:ro1} and Fig.\ref{fig:ro2}.
Similar to the earlier cases of ANNNI, Fig.\ref{fig:ro3} shows on the one hand a decrease of $\xi$ with $\beta$ followed by a singularity in its first derivative at $\beta_D$ and, on the other hand, a smooth and monotonic increase of the range of order $\xi_r$ as $\beta$ ranges across the disorder point $\beta_D$. Even for $\beta >\beta_D$ we see that $\xi$ takes a while to catch up with $\xi_r$ when both the estimates of correlation length merge asymptotically, as expected. Fig.\ref{fig:ro4} is representative of the correlation behaviour in the portions of the short range ordered sub-region for which $K>0$. Here we notice that the standard $\xi$ curve takes a sharp upturn at a point different from the disorder point $\beta_D$. Once again, $\xi_r$ smoothly varies across $\beta_D$.  From both these observation we could also say that $\xi_r$ continues to remain a more consistent estimate of the correlation length than $\xi$ even for a range of temperature beyond the disorder point $\beta_D$, till the two estimates become nearly equal.

 Further, in Figs.\ref{fig:ro3} and \ref{fig:ro4} we draw attention to an interesting concurrence of the curve $\lambda/a$, with $a\sim 5$, with the curve $\xi_r$ for smaller values of $\beta$ until the former begins to rapidly diverge as the temperature nears $\beta_D$. This pattern of behaviour is seen everywhere in the region with short-ranged modulations, with the proportionality constant $a$ ranging from $5$ to $6$. While we are unable to offer any explanation yet for this concurrence, its implication is easier to appreciate. Namely, it suggests that the short-ranged modulations are best observed over the range of temperature for which $\xi_r$ remains proportional to the period of modulation. As the two curves separate and $\lambda$ begins to climb up in the plot, there would be a substantial decay in correlations over a single period  so that the modulations would not be discernible in any statistical sense. Future Monte Carlo studies could be profitably used to test this proposition.

  Admittedly, our definition of $\xi_r$ appears to be an operational one without a rigorous mathematical basis. In cases where the correlation decay assumes a monotonic exponential character $\xi_r$ will of course match with the standard $\xi$ by design. However, where the two disagree the correctness of $\xi_r$ is ultimately judged by whether it gives a reasonable and a better estimate of the correlation length which it does seem to do as demonstrated in the preceding. A further plausibility argument could also be advanced to support our construction of $\xi_r$. Thus, close to the disorder point $\beta_D$, where the wavelength $\lambda$ of modulations is pushed to larger and larger values, the approximation of monotonic exponential decay for the correlation function gets better and better. Here, the range of order $\xi_r$ is quite naturally identified with the correlation length as the point where the (almost) monotonically decaying correlation function drops to $1/e$ times its value of unity. We then continue with this operational definition of $\xi_r$ for lower values of $\beta$ and find that it works well for us.  
  
  Quite remarkably, as we shall soon show, geometry seems to pick $\xi_r$ over the standard $\xi$ as the appropriate estimate of correlation length in the short-ranged modulated order regime. In fact, as we had mentioned in \cite{khatua1}, it was our observations of the state space scalar curvature across the disorder point that had initially motivated the construction of $\xi_r$.

\begin{figure*}[ht]
    \centering
    \begin{subfigure}{0.31\textwidth}
        \centering
        \includegraphics[width=\linewidth]{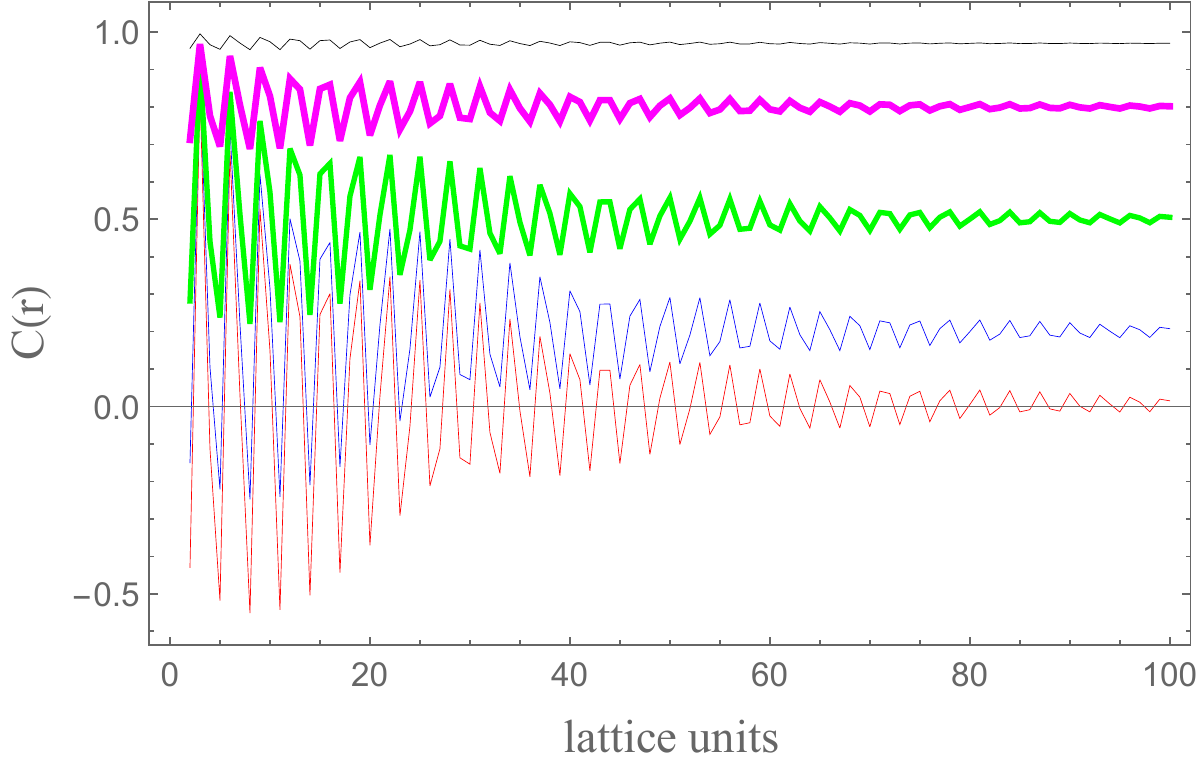}
        \caption{}
        \label{fig:int1}
    \end{subfigure}%
    \hfill
    \begin{subfigure}{0.31\textwidth}
        \centering
        \includegraphics[width=\linewidth]{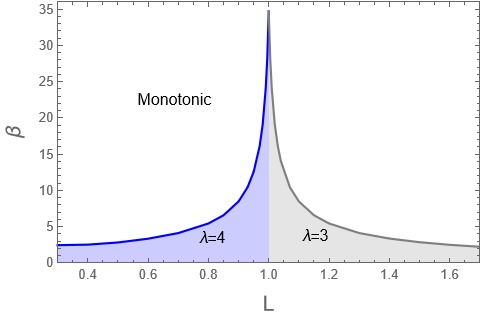}
        \caption{}
        \label{fig:int2}
    \end{subfigure}%
  \hfill
   \begin{subfigure}{0.31\textwidth}
        \centering
        \includegraphics[width=\linewidth]{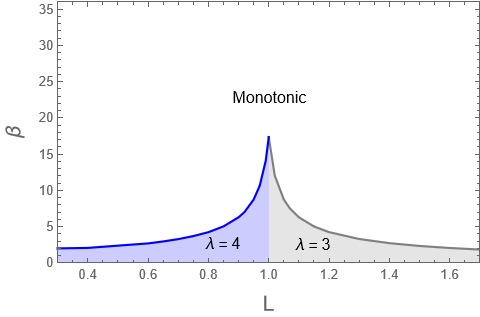}
        \caption{}
        \label{fig:int3}
    \end{subfigure}%
    \caption{\small{Intermediate-ranged modulations: ($a$) show a series of plots of the correlation function $\mathcal{C}(n)$ {\textit{vs.}} lattice units with $\beta=6.27260,6.27281,6.27284$, $6.27287$, and $6 .27295$ from the lowermost to the uppermost curve. The $L$,$K$ values are $1.1$ and $-0.99$, positioned slightly above the $\langle\infty|21\rangle$ boundary in the ferromagnetic phase. ($b$) and ($c$) show plots of transition temperature from oscillatory to monotonic decay.. The plots are taken along lines parallel to the phase boundaries $\langle\infty|2\rangle$ and $\langle\infty|21\rangle$ at normal distances $d=0.005$ and $0.01$ respectively.  }}
    \label{fig intermediate}
\end{figure*}

\subsubsection{Intermediate ranged modulations}

As mentioned in the preceding, spin-spin correlations undergo a slower oscillatory decay at high temperatures in the sub-region where $L>0, K<0$. Since significant modulations persist only a few times the correlation length which is anyway small  at high temperatures we term the modulations as having an intermediate range. Here, unlike the short range case, the wavelength of modulations $\lambda$ remains more or less temperature independent. Moreover the wavelength of modulations follows that of the adjoining region, either $\lambda=4$ for regions adjoining the phase boundary $\langle\infty|2\rangle$ or $\lambda=3$ for regions adjoining $\langle\infty|21\rangle$. Here the `envelope' of oscillatory decay is relevant so that the standard expression in Eq.\ref{corr} gives the correct measure of correlation length.

Interestingly, the change from oscillatory to monotonic decay of correlations here is very different in character from the short range case. Instead of the wavelength diverging to infinity at the disorder point, the correlation function `shifts' upwards over a very narrow range of temperature, thus rendering the decay monotonic. In Fig.\ref{fig:int1} we plot the correlation functions across a very narrow range of $\beta$ values  at a point close to the $\langle 21\rangle$ region. We observe that the oscillations shrink in amplitude as they shift up, thus effectively killing the modulation. Note that the wavelength of modulations remains constant at $\lambda\sim 3$ here.

This also means that while the transition from oscillatory to monotonic decay is not exactly defined as in the previous case, it occurs across a narrow enough temperature range  to remain reasonably sharp for practical purposes. In Fig.\ref{fig:int2} and Fig.\ref{fig:int3} we plot the transition temperature across a range of $K$ values as we move rightwards keeping constant normal distances $d=0.005$ and $d=0.01$ respectively from the phase boundaries $\langle\infty|2\rangle$ and $\langle\infty|21\rangle$. It is seen that the transition  from oscillatory decay to monotonic decay occurs at higher temperatures the further one moves away from the boundaries.

We now briefly recall a few facts about thermodynamic geometry before we set up the state space scalar curvature probe of the A3NNI chain.

 \section{Thermodynamic geometry}  
 \label{sec3}
 
  Riemannian thermodynamic geometry, as pioneered by Weinhold, \cite{wein} and Ruppeiner, \cite{rupporiginal}, has been very successful in elucidating the nature and strength of underlying statistical interactions in diverse physical systems ranging across fluids, spin systems, black holes, strongly interacting matter and several others. A Riemannian manifold structure of the equilibrium state space is obtained by introducing an invariant, positive definite distance measure via the Hessian of the entropy density or, equivalently, that of the Massieu function density
 \begin{eqnarray}
 d\ell^2&=&-\frac{\partial^2s}{\partial x^i\partial x^j}dx^idx^j\nonumber\\
 &=&\frac{\partial^2\psi}{\partial\theta^i\partial\theta^j}d\theta^id\theta^j
 \label{distance measure}
 \end{eqnarray}
 where $\psi=s-x^i\frac{\partial s}{\partial x^i}$ is the Legendre transform of the entropy density. The $x^i$'s are densities while $\theta^i=\partial s/\partial x^i$ are entropic intensive variables.
 The Hessians in Eq.\ref{distance measure} are the state space Riemannian metric $g_{ij}$ expressed in different coordinate systems. We shall find it convenient in this work to use the Massieu function representation of the thermodynamic metric. The Massieu function density is directly related to the partition function $Z$, namely $\psi=\frac{1}{N}\ln Z$ and moreover its Hessian metric has a transparent interpretation as second order fluctuation moments of densities,
  \begin{equation}
  g_{ij}=\frac{\partial^2\psi}{\partial\theta^i\partial\theta^j}=\langle\Delta x_i\Delta x_j\rangle
  \end{equation}

  The Riemannian distance between two nearby states directly informs the probability of equilibrium fluctuations between them. The normalized probability of occurrence of a nearby `fluctuation state' $\bar{x}+dx$ given that the equilibrium state as dictated by the external reservoirs is $\bar{x}$ (standing for $n$ densities $\bar{x_i}$) is given as the following covariant (co-ordinate independent) expression
\begin{equation}
P(\bar{x}+dx)=\frac{1}{(2\pi)^{n/2}}\exp(-\frac{1}{2}d\ell^2)\sqrt{g}d^nx.
\label{probability}
\end{equation} 
where $dl^2=g_{ij}dx^idx^j$.
Therefore, lesser the invariant distance between two equilibrium states greater are the chances of fluctuation between them, making it harder to `distinguish' between the two states. The length of a finite path between two state space points then becomes a measure of the maximum number of distinguishable equilibrium states along the path. 

\subsection*{Thermodynamic curvature: third order moments}

 In general the geometry is curved in the presence of interactions, e.g for the van der Waals gas and is flat or almost flat when interactions are absent, as in the case of the ideal gas. In fact, the scalar curvature $R$ of the thermodynamic manifold turns out to have a deep connection with the underlying statistical interactions in the physical system. This connection is best exemplified by the geometry near a critical point where the invariant $R$ diverges to negative infinity exactly as the inverse of the singular part of the free entropy, modulo an order unity constant that depends only on the universal scaling exponents. Using hyperscaling this also means that $R$ equals the correlation volume near the critical point,
\begin{equation}
R=\kappa_1\psi_s^{-1}=\kappa_2\xi^d
\label{Ruppener conjecture}
\end{equation}
where $\kappa_1,\kappa_2$ are order unity constants that depend only on the universal scaling exponents, \cite{rupprev}.

Furthermore, both in the sub-critical and in the super-critical region $R$ continues to track the correlation length, a feature that has been put to good use in drawing coexistence curves and the Widom line in fluids, spin systems and black holes, \cite{sahay1,may,sarkar1,rupprep}. Remarkably still, $R$ appears to carry clues about the qualitative nature of the underlying interactions. It turns out, the sign of $R$ changes depending on whether the effective interactions are attractive (bosonic, fluid like, critical) or repulsive (fermionic or solid like), \cite{rupprep,behrouz2}. 
Thus, while remaining a purely thermodynamic tool, $R$ is able to probe deep into the mesoscopics of the system.

In the usual case of a two dimensional thermodynamic manifold, as in the present work, the scalar curvature $R$ can be written as a ratio of a third order and a second order determinant, \cite{rupprev},

\begin{equation}
R=\frac{\begin{vmatrix}
\psi_{11}&\psi_{12}&\psi_{22}\\\psi_{111}&\psi_{112}&\psi_{122}\\\psi_{112}&\psi_{122}&\psi_{222}
\end{vmatrix}}{2\begin{vmatrix}
\psi_{11}&\psi_{12}\\ \psi_{21}&\psi_{22}
\end{vmatrix}^{2}}.
\label{ruppdet}
\end{equation}
where the subscripts of $\psi$ represent derivatives with respect to $\theta_1,\theta_2$, which in turn represent fluctuation moments of densities. For instance, $\psi_{21}$=-$\langle\delta \epsilon\delta m\rangle$, $\psi_{122}$=-$\langle\delta m^2\delta\epsilon\rangle$, etc with $m$ and $\epsilon$ being the magnetization and energy per lattice site. 
$R$ therefore appears as a combination of second and third fluctuation moments in the densities. What is significant is that this specific combination of moments is an {\textit{invariant}} one, a geometrical object. Undoubtedly, this motivates a search for a deeper, physical meaning to the peculiar combination of statistical moments presented in Eq.(\ref{ruppdet}) above. 

As has been confirmed by several past works, the success of thermodynamic $R$ lies in the fact that it adds valuable information about mesoscopic fluctuating structures (both quantitative and qualitative) that is not available from standard response functions based on second moments, like the heat capacity $C_H=\beta^2\psi_{11} $ or the susceptibility $\chi_m=\beta\psi_{22}$, etc. Obtained directly via a thermodynamic route, $R$ therefore holds the promise of complementing the analyses from simulations as well as more involved statistical mechanical calculations and calculations based on quantum field theory or even string theory.

Looked at the other way round, the invariant $R$ can also be viewed as a means to interpret some third order moments (and specific combinations thereof) as meaningful descriptors of phase behaviour in thermodynamic systems. 
As it turns out, the expression of $R$ for a reflection symmetric lattice spin systems becomes greatly simplified in the absence of a symmetry breaking magnetic field, thus offering a more transparent connection with higher order moments. With all the fluctuation moments containing odd powers of $\delta m$, namely $\psi_{12},\psi_{211}$, and $\psi_{222}$ vanishing in zero field, the state space curvature simplifies to, \cite{khatua1},
\begin{eqnarray}
R_0&=&\frac{1}{2\,\psi_{11}}\frac{\partial\log\psi_{22}}{\partial\beta}\left(\frac{\partial\log\psi_{11}}{\partial\beta}-\frac{\partial\log\psi_{22}}{\partial\beta}\right)\nonumber\\&=&\frac{1}{2\langle\delta\epsilon^2\rangle}\frac{\langle\delta m^2\delta\epsilon\rangle}{\langle\delta m^2\rangle}\left(\frac{ \langle\delta\epsilon^3\rangle}{\langle\delta\epsilon^2\rangle}-\frac{\langle\delta m^2\delta\epsilon\rangle}{\langle\delta m^2\rangle} \right)\nonumber\\
&=& -\frac{1}{2}\,\frac{\alpha_m\,\rho}{\sigma_\epsilon^2}
\label{curvature zero}
\end{eqnarray}
where we define the third order statistical quantities
\begin{eqnarray}
\alpha_\epsilon &=& -\frac{\partial\log\psi_{11}}{\partial\beta} = \frac{\langle\delta\epsilon^3\rangle}{\langle\delta\epsilon^2\rangle}\nonumber\\
\alpha_m &=& -\frac{\partial\log\psi_{22}}{\partial\beta} = \frac{\langle\delta m^2\delta\epsilon\rangle}{\langle\delta m^2\rangle}\nonumber\\
\rho &=&\alpha_m - \alpha_\epsilon= \frac{\partial\log(\psi_{11}/\psi_{22})}{\partial\beta}
\label{Rzero moments}
\end{eqnarray}
along with the second order fluctuation moments in the densities $\sigma_\epsilon^2 = \psi_{11}=\langle\delta\epsilon^2\rangle$ and $\sigma_m^2 =\psi_{22}=\langle\delta m^2\rangle$. The quantities $\alpha_m$ and $\alpha_\epsilon$ can be understood as the relative changes with temperature in the fluctuation moments in magnetization $m$ and the internal energy $\epsilon$ respectively. The quantity $\rho$ is therefore the difference in the two strain rates, so to speak. As we shall extensively verify in the following, the expression for zero field curvature in Eq.(\ref{curvature zero}) transparently represents the interplay of third order and second order moments in governing the response of geometry.

 \label{sec3sub1}

\section{Geometry of the A3NNI chain}  
 \label{sec4}

In this section we shall endeavour to understand the nature of information revealed by the state space scalar curvature $R$ of the A3NNI chain. Before embarking on our exploration of the state space geometry, however, we make a basic but important observation. 
Namely, as we explain below, the Riemanninan geometry associated with the state space of the A3NNI chain (in zero or non-zero field) is two-dimensional, but not because its intensive parameter space consisting of $\beta$ and $H$ is two-dimensional. 

\subsection{State space dimensionality}

As we had pointed out in \cite{riek1} the correct measure of the number of state space dimensions is the number of {\it{independent}} stochastic variables in the thermodynamic system. Rewriting the A3NNI chain Hamiltonian in Eq.\ref{hamiltonian} as a sum of stochastic variables we get
\begin{eqnarray}
\mathcal{H}&=&\mathcal{F}_1-H\mathcal{F}_2,\,\,\mbox{where}\nonumber\\
\mathcal{F}_2&=& \sum_i S_i,\,\,\,\,\,\,\, \mbox{and}\nonumber\\
\mathcal{F}_1&=&-J_1\sum_i( S_iS_{i+1}+KS_iS_{i+2}+LS_iS_{i+3}).
\label{Hamiltonian stochastic}
\end{eqnarray}
Here $\mathcal{F}_2$ is the  fluctuating magnetization given in terms of the instantaneous configuration of spins while $\mathcal{F}_1$ is the fluctuating internal energy that depends on the relative orientation of the spins. Clearly, the two have no kinematic dependence on each other. The averages of these stochastic variables
\begin{eqnarray}
\frac{1}{N}\langle\mathcal{F}_2\rangle&=&\frac{\partial\psi(\beta,\nu)}{\partial \nu}=m\nonumber\\
\frac{1}{N}\langle\mathcal{F}_1\rangle&=&-\frac{\partial\psi(\beta,\nu)}{\partial \beta}=\epsilon=u+Hm\nonumber\\
\label{stochastic averages}
\end{eqnarray}
are the thermodynamic variables of magnetization per site $m$ and the internal energy per site $\epsilon$.  The variable $u$, obtained as  $u=-\partial\psi(\beta,H)/\partial\beta$, stands for the total energy per site and it includes the interaction of the lattice spin with the external field $H$. The fluctuation moments in the two stochastic variables are obtained via the higher order derivatives of the free entropy or the Massieu function and can be related to the thermodynamic response functions in the standard manner. In zero field even as the parameter space becomes one-dimensional and there is no explicit $\mathcal{F}_1$ term in the Hamiltonian, it still continues to exist as an independent stochastic variable. This is evidenced, for example, by the non-zero fluctuation moments in the magnetization which gives rise to the well known zero-field susceptibility. Also, in the mean field approximation for example,  the Hamiltonian of Eq.\ref{hamiltonian} would reduce to a single site expression, thus rendering $\mathcal{F}_1$ as the only independent stochastic variable. Thus, even though its intensive parameter space shall continue to remain two-dimensional the state space geometry of the mean field A3NNI model would become one-dimensional. The above two examples shall make it clear that the dimensionality of the Riemannian state space depends on the number of statistically independent fluctuating quantities in the thermodynamic system.

\subsection{Thermodynamic curvature of the A3NNI chain}

From Figure \ref{fig phase diagram} for the phase diagram we may classify the phase regions into three classes of ordering according to their thermodynamic behaviour near the ground state: singularly ordered, non-singularly ordered and frustrated. The `singularly ordered regions', consisting of the ferromagnetic and ferri-magnetic regions and some of their associated boundaries, are distinguished by a divergence (singularity) in magnetic fluctuations $\sigma_m^2$ as the ordering increases (and hence $\xi$ diverges) towards zero temperature. The `non-singularly ordered regions', consisting of the anti-ferromagnetic region $\langle 1\rangle$, the anti-phase regions $\langle 2\rangle,\langle 3\rangle$ and some associated boundaries, undergo a decay in magnetic fluctuations $\sigma_m^2$ as $\xi$ diverges towards the ground state. 
Lastly, the `frustrated regions' consist of the three multiphase lines where the short range order and finite entropy per spin in the ground state is characterised by a  small but finite asymptotic values of $\sigma_m^2$ and $\xi$.

We first note that everywhere in the $L$-$K$ parameter space all the moments of the energy fluctuations, namely $\langle\delta\epsilon^2\rangle, \langle\delta\epsilon^3\rangle$, etc decay to zero towards the ground state. The fact that a majority of fluctuations from the mean state (which is closer to the ground state configuration at low temperatures) cost energy implies that even the odd moments remain positive. Further information about energy fluctuations is revealed from asymptotic behaviour of $\alpha_\epsilon$. Thus, we find that asymptotically
\begin{equation}
\alpha_\epsilon=-\frac{\partial\log\psi_{11}}{\partial\beta}=\frac{\langle\delta\epsilon^3\rangle}{\langle\delta\epsilon^2\rangle}\longrightarrow \mathcal{O}(+1),\,\,\,\mbox{everywhere.}
\label{alpha epsilon}
\end{equation}

The quantity $\alpha_m$ on the other hand varies across different regions. We find that, asymptotically,
\begin{equation}
\alpha_m=-\frac{\partial\log\psi_{22}}{\partial\beta}=\frac{\langle\delta m^2\delta\epsilon\rangle}{\langle\delta m^2\rangle}\longrightarrow\begin{cases}
\mathcal{O}(-1) & \mbox{singular}\\ \mathcal{O}(+1) & \mbox{non-singular}\\\pm 0 & \mbox{frustrated}\end{cases}
\label{alpha m}
\end{equation}

The quantity $\alpha_m$ decays to zero on the multiphase lines since, as mentioned above, the denominator $\langle\delta m^2\rangle$ remains finite there even as the numerator goes to zero.

Before surveying the geometry of the three differently ordered regions, we recall the expression for the zero field curvature from Eq.(\ref{curvature zero})
\begin{equation}
R_0=-\frac{1}{2}\frac{\alpha_m \rho}{\sigma_\epsilon^2}\nonumber
\end{equation}
In the ordered regions, singular or non-singular, where $\alpha_m$ remains finite (see Eq.\ref{alpha m}) the expression for zero field $R$ can be further massaged into
\begin{equation}
R_0=\frac{1}{2}\frac{\alpha_m^2}{\sigma_\epsilon^2}\,(\gamma-1)\,\,\,\,\,\,\,\,\,\,\,\,\,\,\,\mbox{(ordered regions)}
\label{R ordered}
\end{equation}
where we define 
\begin{equation}
\gamma=\frac{\alpha_\epsilon}{\alpha_m}
\label{gamma}
\end{equation}

as the ratio of the two `strain rates' $\alpha_m$ and $\alpha_\epsilon$. Interestingly, the ratio exhibits a precise asymptotic (ground state) behaviour. Thus,
\begin{equation}
\gamma\longrightarrow\begin{cases}-1 & \mbox{singular ordering}\\+1 & \mbox{non-singular ordering}\end{cases}
\label{gamma}
\end{equation}
In the following we shall first outline the behaviour of the scalar curvature in the aforementioned three types of regions and then follow it up with a detailed survey of the ground state curvature.

{\textit{The singularly ordered regions}}: In this case we have from Eq.\ref{alpha m}, Eq.\ref{R ordered} and Eq.\ref{gamma} that $R$ diverges to $-\infty$ towards the ground state just as the inverse of the energy fluctuation moment $\sigma_\epsilon^2$. The latter is of the same order as the singular free entropy per spin. In fact, $R_0$ goes  {\it{exactly}} as the inverse of the singular free entropy,

\begin{figure}[!t]
\centering
\includegraphics[width=2.5in,height=2.2in]{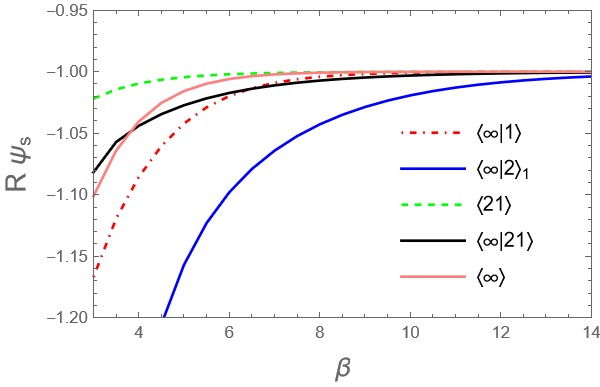}
\caption{\small{The product $R\,\psi_s$ shown converging to $-1$ in all the standard critical regions of ferromagnetic type, wherein the susceptibility diverges to infinity.  }}
\label{fig curvature Rpsi}
\end{figure} 
\begin{figure}[!t]
\centering
\includegraphics[width=2.6in,height=2.2in]{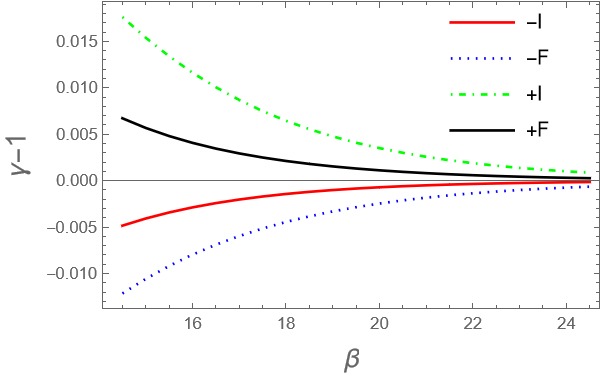}
\caption{\small{ The four distinct ways of approach of $\gamma-1$ to zero.  }}
\label{fig gamma}
\end{figure} 

\begin{equation}
R\,\psi_s\to -1\,\,\,\,\,\,\,\,\,\mbox{          (for critical ordering)}
\label{R divergence attractive}
\end{equation}
thus confirming Ruppeiner's conjecture near the critical point for Ising systems. 

In Fig.\ref{fig curvature Rpsi} we show how $R$ approaches $\psi_s^{-1}$ in all the critical regions. The equality of the scalar curvature with the correlation length can then be checked from Table 1. The divergence of $R$ along with its appropriate scaling behaviour is consistent with its well known presentation near the critical point for several thermodynamic systems, \cite{rupprev,ruppajp}. Its negative signature correctly identifies the attractive nature of effective interactions near the critical region given the presence of long-range correlations and collective spin fluctuations.

{\textit{The non-singularly ordered regions}}: In the case of non-singular ordering it is seen from Eq.(\ref{gamma}) that $\gamma\to 1$ so that the signature and strength of $R$ depends on how fast and from which side $\gamma$ approaches unity. It turns out there are four distinct ways of doing this, as shown in Fig.(\ref{fig gamma}). Thus, $\gamma$ could either increase or decrease to unity resulting in, respectively, negative or positive $R$. In either of the two cases, the numerator $(\gamma-1)$ is seen to approach zero either as fast as the denominator $\sigma_\epsilon^2$ or slower. In the former case $R$ tends to a finite (order unity) positive or negative value while in the latter case it diverges with either sign but at a rate slower than the correlation length.

\begin{figure}[!t]
\centering
\includegraphics[width=3.6in,height=3.6in]{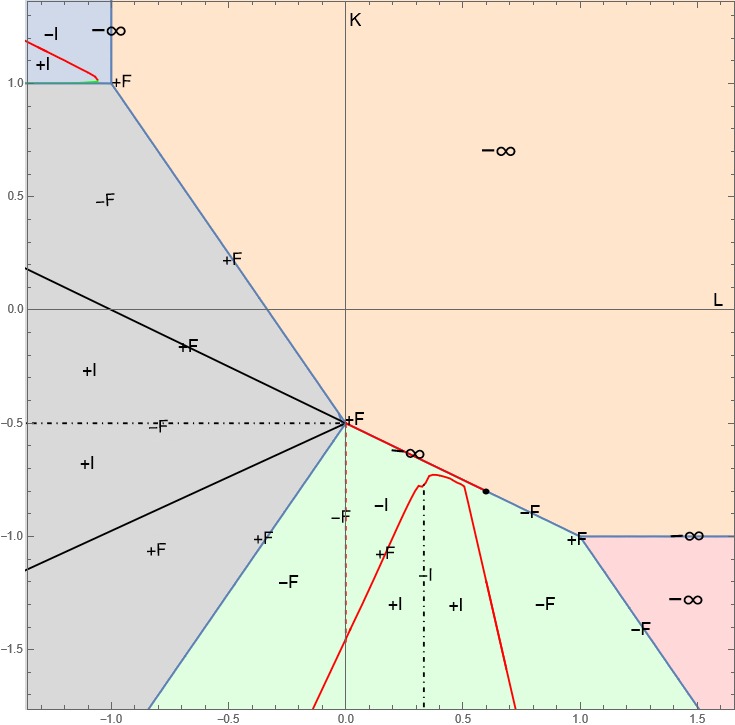}
\caption{\small{Ground state curvature of the A3NNI chain in the $L$-$K$ parameter space. The labels $\pm I$, $\pm F$ and $-\infty$ are explained in the text.}}
\label{fig curvature landscape}
\end{figure} 

\begin{table*}[t!]  
  \centering
 \scalebox{1.1}{
   \begin{tabular}{|c|c|}
\hline \text { Ground state } $R$ &   Effective mesoscopic interactions  \\
\hline\hline$-\infty$ & Critical, collective fluctuations  \\
\hline $+F$ & Weakly repulsive\\
\hline $-F$ & Weakly attractive\\
\hline $+I$ & Strongly repulsive\\
\hline $-I$ & Strongly attractive\\
\hline  
\end{tabular}
}
  \caption{\small{Table listing different categories of ground state zero field curvature and the effective interactions they represent. }}
  \label{tab R label}
\end{table*}

{\textit{The frustrated regions}}: As mentioned earlier, the boundaries between ground state phases $\langle \infty\rangle$ and $\langle 3\rangle$, between $\langle 3\rangle$ and $\langle 2\rangle$ and between $\langle 2\rangle$ and $\langle 21\rangle$ are the multiphase lines. These are the bold boundary lines in Fig.\ref{fig phase diagram}. From Eq.\ref{alpha m} one sees that along these lines the quantity $\alpha_m$ goes to zero so that the expression for zero field curvature in Eq.\ref{R ordered} is not valid here. Instead, one obtains a simple asymptotic expression for of $R_0$ for the multiphase lines as,
\begin{equation}
R_0 \to \frac{1}{2}\frac{\alpha_m\alpha_\epsilon}{\sigma_\epsilon^2}\,\,\,\,\,\,\,\mbox{(multiphase lines)}
\label{R disordred}
\end{equation}
Along the multiphase lines it is $\alpha_m$ approaching zero at the {\it{same}} rate as $\sigma_\epsilon^2$ that causes a flat asymptotic curvature. Moreover, whether $\alpha_m$ approaches zero from positive or negative values shall determine the sign of curvature.

\begin{figure}[!t]
\centering

\includegraphics[width=3in,height=2.1in]{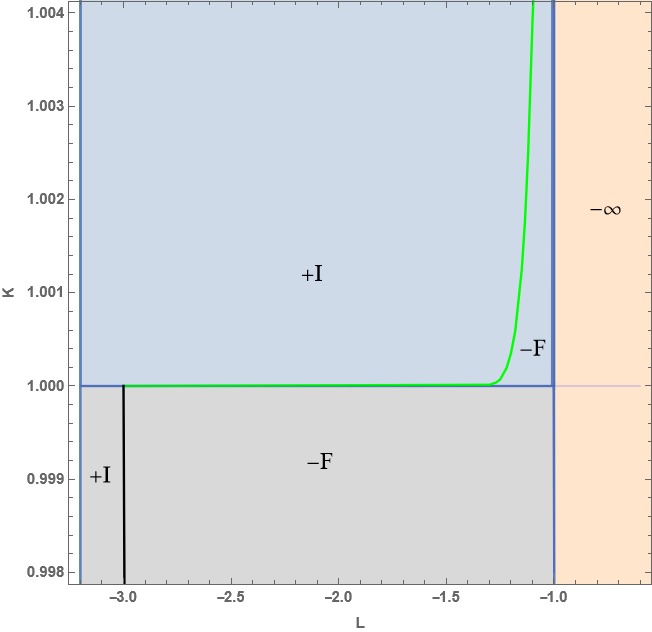}
\caption{\small{A zoom in of the the lower portion of the antiferromagnetic parameter space from Fig.\ref{fig curvature landscape}. The portion under the blue curve has a negative flat asymptotic curvature. Shown in the figure, the blue curve continues leftwards and ends on the $K=1$ line at $L=-3$.   }}
\label{fig curvature landscape A}
\end{figure} 

\begin{figure*}[htbp]
    \centering
 
    \begin{subfigure}{0.3\textwidth}
        \centering
        \includegraphics[width=\linewidth]{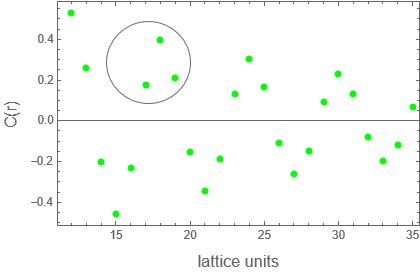}
        \caption{}
        \label{fig:sub1}
    \end{subfigure}%
    \hfill
    \begin{subfigure}{0.3\textwidth}
        \centering
        \includegraphics[width=\linewidth]{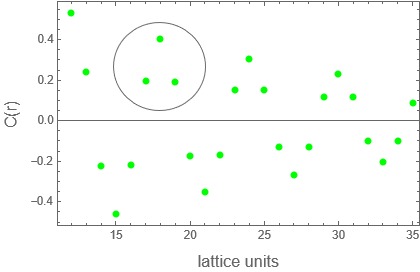}
        \caption{}
        \label{fig:sub2}
    \end{subfigure}%
    \hfill
    \begin{subfigure}{0.3\textwidth}
        \centering
        \includegraphics[width=\linewidth]{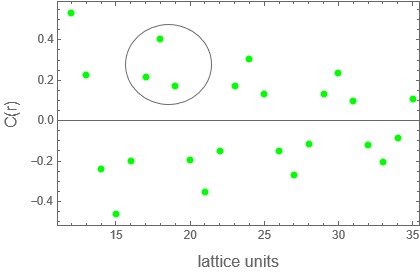}
        \caption{}
        \label{fig:sub3}
    \end{subfigure}\\%
    \begin{subfigure}{0.3\textwidth}
        \centering
        \includegraphics[width=\linewidth]{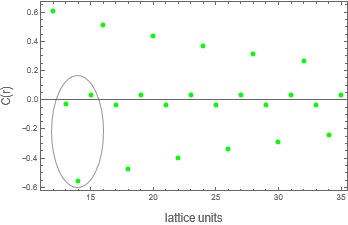}
        \caption{}
        \label{fig:sub4}
    \end{subfigure}%
    \hfill
    \begin{subfigure}{0.3\textwidth}
        \centering
        \includegraphics[width=\linewidth]{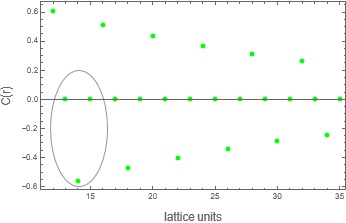}
        \caption{}
        \label{fig:sub5}
    \end{subfigure}%
    \hfill
    \begin{subfigure}{0.3\textwidth}
        \centering
        \includegraphics[width=\linewidth]{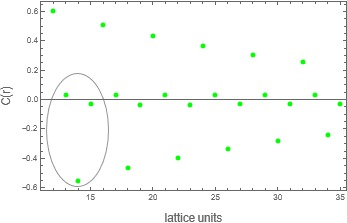}
        \caption{}
        \label{fig:sub6}
    \end{subfigure}
  
    \caption{Plots ($a$),($b$),($c$) with parameters $\beta=1.02,L=-1,K=-0.47,-0.5,-0.53$   show the change in shape of the correlation function across the symmetry line (see text) in the $\langle 3\rangle$ region. Plots ($d$),($e$),($f$) with parameters $\beta=1.02,L=11/30,10/30,9/30,K=-1.6$ show the change in shape of the correlation function across the symmetry line (see text) in the $\langle 2\rangle$ region. Plots ($b$) and ($e$) refer to the symmetry lines in the respective regions. }
    \label{fig corr skew}
\end{figure*}

 \begin{figure*}[htbp]
    \centering
    \begin{subfigure}{0.4\textwidth}
        \centering
        \includegraphics[width=\linewidth]{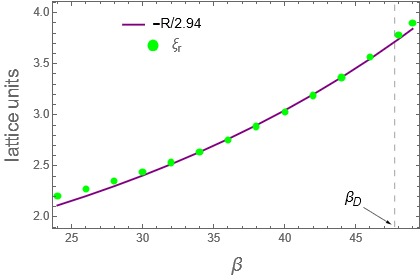}
        \caption{}
        \label{fig:sho1}
    \end{subfigure}%
    \hfill
    \begin{subfigure}{0.4\textwidth}
        \centering
        \includegraphics[width=\linewidth]{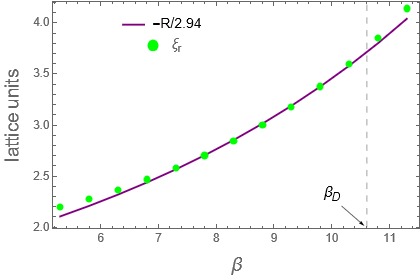}
        \caption{}
        \label{fig:sho2}
    \end{subfigure}%
  
    \caption{\small{($a$) and ($b$) show plots of $R$ and the range of order $\xi_r$ with the same parameter values as in Fig.\ref{fig:ro1} and in Fig.\ref{fig:ro2} respectively. For both the cases  $R$ shows an excellent concurrence with the range of order $\xi_r$ within the modulated order region.   }}
    \label{fig R short}
\end{figure*}

Following our discussion on the general structure of $R$ in the three types of regions we now undertake a more detailed survey of the geometry. In Fig.\ref{fig curvature landscape} we depict in the $L$-$K$ parameter space the asymptotic trends in the state space scalar curvature $R$ in different regions and their adjoining boundaries. The curvature $R$ has been calculated from Eq.(\ref{curvature zero}) and has been evaluated numerically at all points. Four labels mark the ground state curvature: $\pm F$ indicates that $R$ tends to small (order unity) positive or negative values on approaching zero temperature, $-\infty$ means $R$ diverges to minus infinity at the same rate as the correlation length $\xi$ (or $\psi_s^{-1}$), and finally $\pm I$ again indicates a positive or negative divergence (but always at a rate slower than the correlation length).

First, we confirm that the ground state $R$ is indeed $-\infty$ in {\it{all}} of the critically ordered cases which includes the regions $\langle\infty\rangle$, $\langle 21\rangle$ and the line segments $\langle\infty|1\rangle$, $\langle\infty|2\rangle_1$ and $\langle \infty|21\rangle$. Next, we see that the antiphase regions $\langle1\rangle$, $\langle 2\rangle$ and $\langle 3 \rangle$ show a clear demarcation into sub-regions of flat and diverging curvatures with both signatures. 

 As discussed earlier, a growing body of work demonstrates that the signature of $R$ discriminates between the attractive and repulsive nature of underlying statistical interactions. We too shall try to interpret our observations of $R$ in this light. In Table \ref{tab R label} we enumerate the different types of asymptotic trends in $R$ along with the underlying statistical interactions suggested by the trends.

The antiferromagnetic or the $\langle 1\rangle$ region divides into sub-regions with strongly attractive, strongly repulsive and weakly attractive effective interactions, the last region being much smaller than others. It is seen that the strong repulsive interactions occur in the lower half (or the left half) of the $\langle 1\rangle$ region where the negative $L$ coupling is, in a sense, stronger than the positive $K$ coupling.  In Fig.\ref{fig curvature landscape A} we magnify the lower portion of the antiferromagnetic region to reveal a weakly attractive region adjoining its boundary with the ferromagnetic region.

 Interestingly, this is the first instance of an antiferromagnetic ground state showing strong effective interactions, either attractive or repulsive. In earlier cases of antiferromagnetic ordering, whether in the Ising model or in more complex decorated spin model, the curvature $R$ has always remained order unity, \cite{rupprev,belucci}.  We also recall that the $\langle 1\rangle$ region does not maintain the hyperscaling relation between the singular free entropy $\psi_{s}$ and the correlation length $\xi$.
 
While there are some noticeable differences between the two, variations in the effective interactions in the $\langle 3\rangle$ and $\langle 2\rangle$ antiphase regions exhibit broad similarities. Thus we see that in both these regions there is a strongly repulsive core  flanked by sub-regions of weak attraction or repulsion. And much like the $\langle 1\rangle$ region the strongly repulsive core of $\langle 2 \rangle$ and $\langle 3 \rangle$ indicates a relative dominance of, respectively, the negative $K$ and the negative $L$ spin couplings. The repulsive core of both these regions have a central line of `symmetry' along which the effective interactions are attractive, either strong as in $\langle 3\rangle$ or weak, as in $\langle 2\rangle$. For the $\langle 2\rangle$ region the line is the same (though it does not reach all the way up to the boundary $\langle\infty|2\rangle$) as the dashed red line at $L=1/3$ in Fig.\ref{fig phase diagram} across which the critical scaling behaviour changes. No such change in scaling behaviour occurs across the symmetry line $K=-1$ in the $\langle 3\rangle$ region. 

Interestingly, we observe a common pattern of variation in the correlation function across both the lines of symmetry. In Figs.\ref{fig:sub1}-\ref{fig:sub3} we plot correlation function at a fixed temperature for points just above, below and on the symmetry line in $\langle 3\rangle$. And in Figs.\ref{fig:sub4}-\ref{fig:sub6} we do the same for points just to the left, right and on the symmetry line in $\langle 2 \rangle$.  For both the lines we notice a left-right or a top-down skew in the spatial variation of correlations which disappears on the symmetry line. Admittedly, beyond the reasonable supposition that a variation in the {\it{shape}} of the two-point correlation function could possibly be a consequence of a change in  higher point correlations which could in turn be reflected in the state space geometry via the third order moments  we are unable, at present, to make a specific connection between the pattern of variations in the correlation function and those in the ground state $R$. We hope that a future investigation could explain this interesting observation. A notable difference between the $\langle 3 \rangle$ and the $\langle 2\rangle$ phase is the presence of a strongly attractive triangular region in the latter.

Finally, we return to the short-ranged modulated order region of the ferromagnetic phase $\langle\infty\rangle$ discussed in section \ref{sub1sub1} and view it in the light of geometry. We quickly recall that  for this region, which includes all of the ferromagnetic phase with $L$ negative (see Fig.\ref{fig contour plot disorder lines}), we had argued that the `range of order' $\xi_r$ is a better estimate of the correlation length than the standard expression as a logarithm of the ratio of eigenvalues of the transfer matrix. Here we find that geometry nicely corroborates our proposal. Thus, we see that in Fig.\ref{fig:sho1} and Fig.\ref{fig:sho2} (with the same parameter values as in Fig.\ref{fig:ro1} and \ref{fig:ro2}) the range of order $\xi_r$ is very nearly equal to $-R/3$ all the way down to almost about two lattice units. The curvature $R$ and $\xi_r$ vary smoothly across the disorder point $\beta_D$ and for large $\beta$ attain the asymptotic relation $\xi_r=\xi=-R/2$. 

It is remarkable indeed that geometry is able to pick out a certain operational measure of correlation length as the correct one. 

\section{Conclusions}
\label{sec5}

In this work we expand the scope of [1] with our investigations into the phase behaviour and geometry of the A3NNI model, which includes the ANNNI model as a special case. We uncover interesting behaviour like hyperscaling violation and intermediate ranged modulations in the ground state. With two independent couplings $K$ and $L$ the A3NNI model allows for a rich and varied phase behaviour. Our construction of the two-point correlation function is crucial to understanding both short and intermediate range modulations in the ground state. 

Useful as the two-point correlation function is in differentiating ground state spin configurations, nature of modulations, etc there do remain effects that are attributable to higher point correlations and hence are not effectively probed by the spin-spin correlation function or the standard response functions like the susceptibility or the heat capacity. The state space scalar curvature $R$ comes in handy in this regard. While $R$ is a macroscopic thermodynamic quantity unlike the microscopic two-point correlation function it nevertheless contains contributions from higher point correlations. This is because $R$ contains second as well as third order moments in an invariant combination and the higher moments get significant contributions from the higher point correlations. 

It turns out, the geometric description of the ground state involves a further sub-division of its phase diagram into regions of repulsive or attractive effective interactions, both strong and weak.  Our observations of the zero field curvature suggest that the association between the type of spin configuration and the nature of effective interactions is not so direct as implicitly believed earlier. Rather, it appears that the interplay of short range forces brings in nuances to the statistical nature of underlying interactions. That is, tuning the relative strengths of the higher order couplings $J_2$ and $J_3$  effects variations in the underlying mesoscopics of the system which, though not detectable at the level of ground state spin configurations or even from the standard response functions is, remarkably enough, encoded into the state space geometry.
 
 A future Monte Carlo study of frustrated systems could help illuminate such effective interaction mediated meso-structures that are suggested by the geometric description of physical systems. 
 
 We sincerely hope that our work would go some way in motivating and aiding in the effective use of the geometric probe for investigating  thermodynamic systems.

\begin{acknowledgments}
We thank Paras Poswal and Subrat Das for discussions and for help with some figures. 
\end{acknowledgments}
\appendix*
\section*{Appendix}
\label{appen}
\renewcommand{\theequation}{A-\arabic{equation}}
\setcounter{equation}{0}

The nearest neighbour Hamiltonian of the $1d$ Ising model
\begin{equation}
\mathcal{H}=-J\sum_i S_i\,S_{i+1}-H\sum_i S_i
\label{ising model}
\end{equation}
supports a ferromagnetic or an anti-ferromagnetic ground state depending on whether the exchange coupling $J$ is positive or negative.
Using the transfer matrix method with periodic boundary conditions the partition function $Z$ is obtained via the trace of $N$ copies of the transfer matrix  $\hat{\mathcal{T}}$ whose components are expressed as
\begin{equation}
\langle u|{\hat{\mathcal{T}}}|v\rangle = \exp\left[\beta\left( J u\,v+ \frac{H(u+v)}{2}\right)\right],
\label{transfer matrix ising}
\end{equation} 
where the spin variables $u,v$ are $\pm 1$.
We may denote the eigenvectors of the transfer matrix as $|t_1\rangle$, $|t_2\rangle$ and the eigenvalues $\lambda_1>\lambda_2$.
The free entropy (or the Massieu function) per spin becomes the logarithm of the largest eigenvalue in the limit of large $N$
\begin{equation}
\psi(\beta,\nu)=\log \lambda_1
\label{ising free energy}
\end{equation}
where the entropic intensive variables are $\beta=1/kT$ and $\nu=H\beta$. In terms of the above quantities the correlation function becomes
\begin{equation}
\langle S_iS_{i+r}\rangle = \langle t_1|{\hat{\mathcal{S}}}|t_1\rangle^2 + \left(\frac{\lambda_2}{\lambda_1}\right)^rf(T)
\label{ising correlation fn 1}
\end{equation}
where $f(T)=\langle t_1|\hat{\mathcal{S}}|t_2\rangle\langle t_2|\hat{\mathcal{S}}|t_1\rangle$,  $\delta S_i=S_i-\langle S_i\rangle$, the spin matrix $\hat{\mathcal{S}}=\text{diag}[1,-1]$ and $\langle t_1|\hat{\mathcal{S}}|t_1\rangle$ is the average spin per lattice site. Due to translation invariance Eq.(\ref{ising correlation fn 1}) above can also be written as a correlation of spin fluctuations,
\begin{equation}
\langle \delta S_i\delta S_{i+r}\rangle = (\text{sgn}\,\lambda_2)^r \exp\left(-\frac{r}{\xi}\right)f(T)
\label{ising correlation fn 2}
\end{equation}
where the correlation length
$\xi^{-1}=\log \lambda_1/|\lambda_2|$
approaches infinity if $|\lambda_2|\to \lambda_1$.  It can be checked that $f(T)=1$ for both signs of $J$ when $H=0$. 
Meanwhile, the mean square fluctuation in the magnetization  $\sigma_m^2=\partial^2\psi/\partial\nu^2$ can be expressed as a sum over fluctuation correlations across all length scales
\begin{eqnarray}
\sigma_m^2&=&\sum_{i=0}^N \langle \delta S_1\delta S_{1+i}\rangle \nonumber\\
&=& 2\sum_{i=0}^{N/2} \langle \delta S_1\delta S_{1+i}\rangle- \langle \delta S_1\delta S_{1}\rangle\nonumber\\
&=& f(T) \frac{1+\lambda_2/\lambda_1}{1-\lambda_2/\lambda_1}
\end{eqnarray}
where the second equality follows from the fact that in a periodically identified, translational invariant lattice the correlations repeat as 
\begin{equation}
\langle S_1S_{1+r}\rangle=\langle S_1S_{N+1-r}\rangle
\end{equation}
assuming $N$ to be even. Thus the correlation length can cover at most half the spins.

For the ferromagnetic case ($J>0,H=0$) where the second eigenvalue $\lambda_2$ is positive we find that the variance becomes
\begin{equation}
\sigma_m^2=\frac{1+e^{1/\xi}}{1-e^{1/\xi}}\to 2\xi\,\,\,\,\text{(fm)}
\end{equation}
On the other hand for the zero field anti-ferromagnetic case $(J<0, H=0)$ where the second eigenvalue $\lambda_2$ becomes negative it follows that the fluctuation moment decays as the inverse of correlation length,
\begin{equation}
\sigma_m^2= \frac{1-e^{1/\xi}}{1+e^{1/\xi}}\to \frac{1}{2\xi}\,\,\,\,\text{(afm)}
\end{equation}
Therefore, in the anti-ferromagnetic case the variance $\sigma_m^2$ informs in a precise manner the growth of anti-parallel ordering as the temperature is lowered.
\twocolumngrid

\end{document}